\documentclass[final,5p,times,twocolumn,unknownkeysallowed]{elsarticle}

\usepackage{doi}
\usepackage{hyperref}
\hypersetup{
  colorlinks=true,        
  linkcolor=blue,         
  citecolor=cyan,         
}

\usepackage{graphicx}
\usepackage{dcolumn}
\usepackage{bm}
\usepackage{color}
\usepackage{amsmath}
\usepackage{amssymb}

\begin{document}
\renewcommand{\baselinestretch}{1.3}

\title{Rotating charged black hole in $4D$ Einstein-Gauss-Bonnet gravity: Photon motion and its shadow}

\author[mainaddress1,mainaddress2]
{Uma Papnoi}
%
\author[mainaddress3,mainaddress4,mainaddress5,mainaddress6]
{Farruh~Atamurotov\corref{cor2}}\cortext[cor2]{Corresponding author}
\ead{atamurotov@yahoo.com}

\address[mainaddress1]{Gurukul Kangri (Deemed to be University), Haridwar 249404, Uttarakhand, India}
\address[mainaddress2]{Kanoria PG Mahila Mahavidyalaya, Jaipur 302004, Rajasthan, India}
\address[mainaddress3]{Inha University in Tashkent, Ziyolilar 9, Tashkent, 100170, Uzbekistan}
\address[mainaddress4]{Akfa University, Kichik Halqa Yuli Street 17,  Tashkent 100095, Uzbekistan}
\address[mainaddress5]{Department of Astronomy and Astrophysics, National University of Uzbekistan,Tashkent  100174, Uzbekistan}
\address[mainaddress6]{Institute of Nuclear Physics, Tashkent 100214, Uzbekistan}

\date{\today}
\begin{abstract}
 We construct a charged rotating black hole in 4D Einstein-Gauss-Bonnet (EGB) gravity starting from charged black hole in 4D EGB gravity using complex coordinate transformations suggested by Newman–Janis. Further, we have studied the null geodesics to investigate the shape of the shadow cast by a rotating charged black hole in 4D EGB  gravity. Also we have discussed their horizon properties and shadow cast. The phenomenon of black hole shadows alongwith the horizon structure and energy emission has been analysed to see the influence of Gauss-Bonnet term and black hole charge on horizon, shadow, effective potential and energy emission rate which are compared to their non rotating counterpart. It has been seen that the Gauss Bonnet parameter have an influence on the shape and size of the shadow as well as on the effective potential, horizon and energy emission rate. 
\end{abstract}



\maketitle

\section{Introduction}
General relativity (GR) is an elegant theory which agrees with all observations at solar system scale and beyond. Black holes (BHs) are natural candidates to investigate strong curvature corrections to GR. Recent astrophysical observations and experiments related to the gravitational waves  and the first image of the supermassive BH shadows by Event Horizon Telescope (EHT) Collaboration released located at the center of M87 galaxy \cite{PhysRevLett.116.061102,Abbott_2016,Akiyama19L1,Akiyama19L6,Akiyama19L2,Akiyama19L3,Akiyama19L4,Akiyama19L5} allow us to understand the nature of the geometry and also to test the strong gravity regime. The future observations, like the Next Generation Very Large Array \cite{2015IAUGA..2255106H}, the Thirty Meter Telescope \cite{2013JApA...34...81S}, and the Black Hole Cam \cite{Goddi_2017}, provides a good opportunity to look into the regime of strong gravity,
and to distinguish different modified gravities. Although, Einstein theory has the limits of applicability where it can lose its predictive power. Thus, for its validity and applicability higher order theories extensions in GR has been proposed \cite{2012PhLB..711..196D}. In this context, Lovelock theory, whose action is a homogeneous polynomial in Riemann curvature,
is the most natural higher dimensional generalization of
the Einstein gravity \cite{Lovelock:1971yv}. It is worth to note that despite the action being polynomial in Riemann curvature, its equation of motion (EOM) remains second order. Gauss-Bonnet (GB)/Lovelock gravity with the quadratic order gives contribution to the EOM only in
$D>4$. However, it  is possible to make higher order terms contribute
in the equation in 4D by dilaton coupling \cite{Kanti_1996,Maeda_2009,Sotiriou_2014,Kanti_2015}. Recently it is proposed that the novel
Einstein-Gauss-Bonnet (EGB) theory could exist even in
$D = 4$ dimensions without dilaton coupling, thus allowing to avoid the Lovelock’s theorem by rescaling the GB coupling constant \cite{Glavan_2020}. The Gauss-Bonnet action
does not contribute to the dynamics of the four-dimensional
spacetime, as its contribution to Einstein’s equation
vanishes identically in 4D. It was shown that by rescaling the GB coupling constant as $\alpha -> \alpha / (D-4)$ the GB invariant, in the limit $D->4$ finding equations, makes a non-trivial contribution to the gravitational dynamics even in $D=4$ \cite{Glavan_2020}, it results in the discovery of novel 4D static and spherically symmetric BH solution. 

The purpose is to introduce a divergence that exactly cancels the vanishing contribution that the GB term makes to the field equations in 4D. A similar type reduction is seen by Mann and Ross \cite{Mann_1993} for $D=2$. Apart from the usual GR, in 4D EGB gravity the causal structure departs from its counterpart that the region around singularity becomes time-like, which is spacelike in Einstein gravity \cite{dadhich2020causal}. Since in 4D EGB, there are some arguments regarding the redefinition of the GB term and the action for new theory \cite{Hennigar_2020,Arrechea_2020,G_rses_2020,Mahapatra_2020}, still the new 4D EGB theory as an alternative to Einstein's theory attracted strong attention. In this context, recently several work has been done to understand the nature of EGB theory in 4D \cite{Liu_2021,Guo_2020,wei2020testing,Kumar_2020,Konoplya_2020,churilova2020quasinormal,Malafarina:2020pvl,Aragon:2021ogo,Hosseini_Mansoori_2021,Ge:2020tid,Rayimbaev:2020lmz,Chakraborty:2020ifg,odintsov2020rectifying,Odintsov:2020zkl,Lin:2020kqe,Aoki_2020,Shaymatov:2020yte,Islam_2020,SINGH2020135658,Yang2020bb}. Very recently much of the analyses involved the impact of GB term on the superradiance \cite{zhang2020superradiance}, the motion of spinning particle \cite{zhang2020spinning}, the scalar and electromagnetic perturbations in testing the strong
cosmic censorship conjecture \cite{Mishra_2020} and charged particle
and epicyclic motions around 4D EGB BH \cite{Shaymatov:2020yte}. Later,
following \cite{Glavan_2020} the charged \cite{Fernandes_2020,2021arXiv210502214A} and rotating \cite{Kumar_2020} analogues were obtained for new 4D EGB theory. It is worth noting
that some properties of GB BH in higher dimensions were also investigated in Refs. \cite{Abdujabbarov:2015rqa,Aguilar-Perez:2019vhj,Dadhich:2021vdd,Konoplya20b}.
Further, the EHT observation has restimulated the study of the BH shadow. BH shadow has also been widely analyzed in different gravity models. Synge \cite{synge1960classical} and Luminet in \cite{Luminet79a}
describe the shadow size. Several papers describing the shape of BH shadows with various parameters in various theories of gravity have been investigated in \cite{Luminet79a,Falcke00a,Bambi09a,Amarilla10a,Hioki09a,Abdujabbarov13a,Amarilla12a,Amarilla13a,Abdujabbarov16a,Atamurotov:2013sca,Tsukamoto18a,Atamurotov:2013dpa,Perlick18a,Hou_2018,Cunha20a,2021MNRAS.tmp.1223A,2021PhRvD.103h4005B,2021EPJP..136..436W,2020PhRvD.102j4032G,2021PhRvD.103j4047K,2020EPJC...80.1195H,2000CQGra..17..123D,2015MNRAS.454.2423A,2014PhRvD..89l4004G}. Moreover,  \cite{Atamurotov:2013dpa,Atamurotov:2013sca,Abdujabbarov13a,Cunha20a,Atamurotov:2015nra,Atamurotov:2015xfa,Papnoi:2014aaa,Babar:2020txt,Cunha_2017,atamurotov2021shadow,Frion21,Brahma21}. With the recent observations evidence for the presence of BHs at the centers of galaxies, the study of astrophysical processes in plasma medium surrounding BH has become very interesting. Shadow and gravitational lensing of BH spacetimes in vaccuum has been extended to the study in inhomogeneous and homogeneous plasma around BHs recently (see, e.g. \cite{Perlick15a, Perlick17a,chowdhuri2020shadow,Atamurotov:2015nra, Atamurotov21axion,Babar:2020txt,Atamurotov21PFDM,Javier21,Fathi21a,Schee:2017hof,,Babar2021bbba,Babar2021aaaa,Atamurotov2021aaaa,Atamurotov2021bbba}). The influence of a plasma on the shadow of a spherically symmetric BH is investigated by Perlick et al. \cite{Perlick15a}. With these insights it is also expected the study of BH shadow cast deep insight into modified gravities. In this paper, our main focus is on studying the shadow cast by the charged rotating BH in 4D EGB gravity. Starting from the charged BH in 4D EGB we use Newman Janis to obtain its rotating counterpart by using Newman-Janis-Algorithm (NJA) \cite{Newman:1965tw,Newman:1965my}. 
Furthermore, we look into how the GB coupling constant, rotation parameter and BH charge affect the photon radius, horizon structure, effective potential and  energy emission.

The paper is organized as follows: In Sec. \ref{foursbh}, we have discussed the
charged BH solution in 4D EGB gravity and obtained its rotating counterpart using the NJA. We also discussed the horizon and geodesic structure of the charged rotating BH in 4D EGB gravity in this section. In Sec. \ref{observa}, we have presented the particle motion around the BH solution to discuss the shadow of the BH with its graphical representation by defining the observables to show the effect of GB coupling constant, BH charge and spin parameter. The Emission energy from the BH is investigated in Sec. \ref{emission}. Finally the results have been concluded in Sec~\ref{Conclusions}.
 We have used units that fix the speed of light and the gravitational constant via $G = c = 1$.
\section{Charged 4D Einstein-Gauss-Bonnet black hole}
\label{foursbh}
In this section, we review the metric for charged BH in $4D$ EGB gravity. We start from the action of the Einstein-Maxwell-Gauss-Bonnet theory in 4D spacetime introduced in \cite{Glavan_2020, Fernandes_2020} after rescaling the Gauss-Bonnet (GB) coupling constant. The gravitational action of the EGB theory with electromagnetic field in $D$ dimensional spacetime as:
\begin{eqnarray}\label{action}
 \mathcal{I} = \frac{1}{16\pi} \int{d^Dx \sqrt{-g}[R + \frac{\alpha}{D-4} {\cal G} -F_{ab}F^{ab}]},
\end{eqnarray} with \begin{eqnarray}
 {\cal G} &=& R_{abcd}R^{abcd}- 4 R_{ab}R^{ab}+ R^2,\nonumber \\
 F_{ab}&=&\partial_a A_b -\partial_b A_a,
\end{eqnarray} are the Gauss-Bonnet term and electromagnetic tensor respectively. Here, $\alpha$ is a finite non-vanishing dimensionless Gauss-Bonnet coupling constant \cite{Fernandes_2020} having dimensions of $[length]^2$ which represents the corrections to Einstein theory, $g$ is the determinant of $g_{ab}$, $A_a$ is the electromagnetic four potential and $R$ is the Ricci scalar.

\subsection{Non-rotating black hole}
\label{nospin}

Considering the limit $D\rightarrow 4$, the charged EGB BH solution of action (\ref{action}) was obtained as 
  
\begin{eqnarray}
 ds^2&=&f(r)dt^2-\frac{dr^2}{f(r)}-r^2(d\theta^2+\sin^2\theta d\phi^2),\label{GBme}\\
 f(r)&=&1+\frac{r^2}{2\alpha}\left[1 \pm \sqrt{1+4\alpha\left(\frac{2 M}{r^3}-\frac{Q^2}{r^4}\right)}\right],\label{meme}
\end{eqnarray} with $M$ and $Q$ is the mass and charge of the BH respectively.  The sign $\pm$ refers to the two different branches of solutions. But since, the negative (-ve) branch is connected to standard Einstein–Hilbert gravity, as it reduces to the general
relativity solution in the limit $\alpha \rightarrow 0$. So, we have considered here the negative solution\cite{Glavan_2020,2021arXiv210502214A}.\\
In the limit $\alpha \rightarrow 0$, Eq. (\ref{meme}), takes the form 
\begin{equation}
   \lim_{\alpha\rightarrow 0} f(r)= 1-\frac{2M}{r}+\frac{Q^2}{r^2}+\frac{4M^2}{r^4}\alpha+...
\end{equation}
one can recover the Reissner-Nordstr$\Ddot{o}$m solution. 

To discuss the radius of the photon sphere, we use the condition 
\begin{eqnarray}\label{photon equation}
 \frac{d}{dr}\Big(\gamma(r)^2\Big)\bigg|_{r=r_{ph}}=0,
\end{eqnarray}
with
\begin{eqnarray}
 \gamma(r)^2=\frac{g_{22}}{g_{00}}=\frac{r^2}{f(r)},
\end{eqnarray}
where $g_{00}$ and $g_{22}$ are elements of metric.

\begin{figure*}
 \begin{center}
   \includegraphics[scale=0.7]{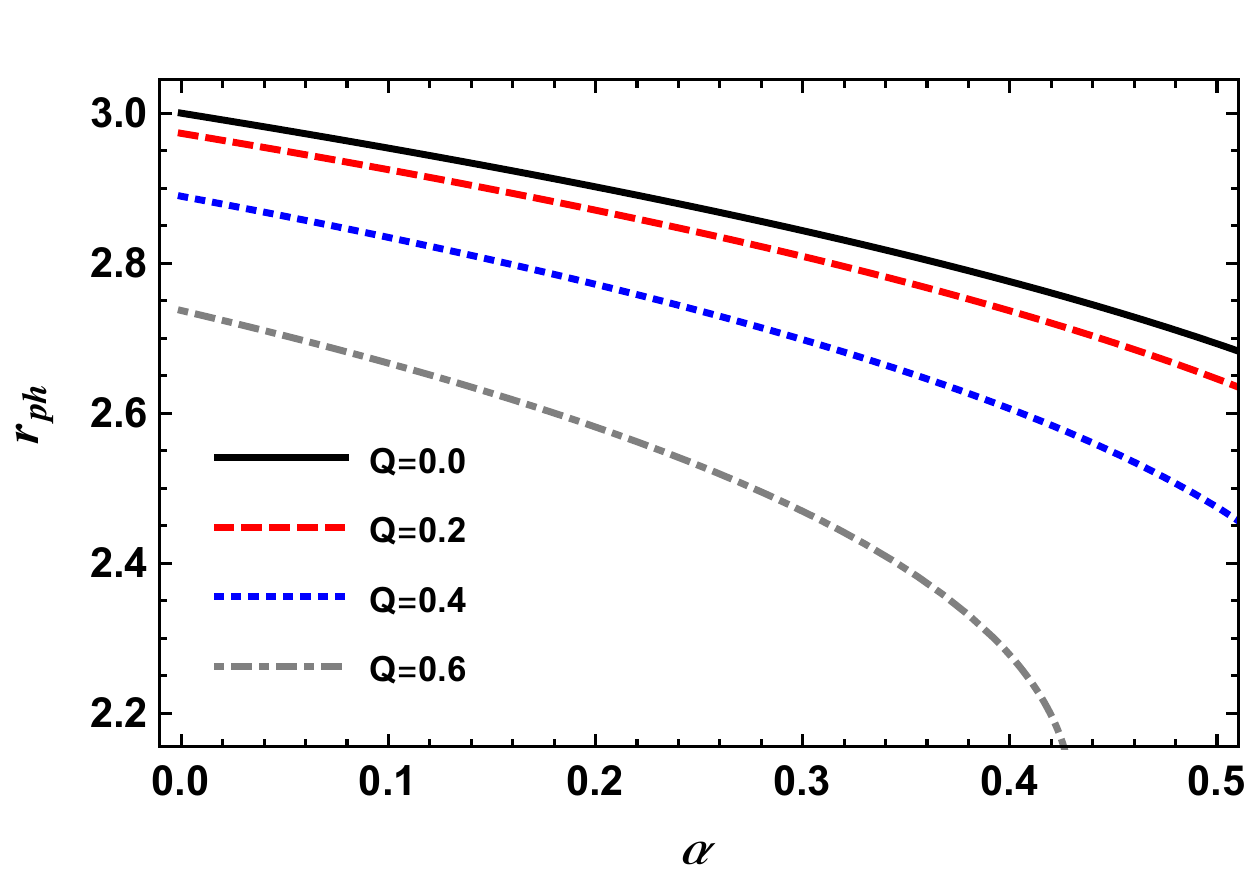}
   \includegraphics[scale=0.7]{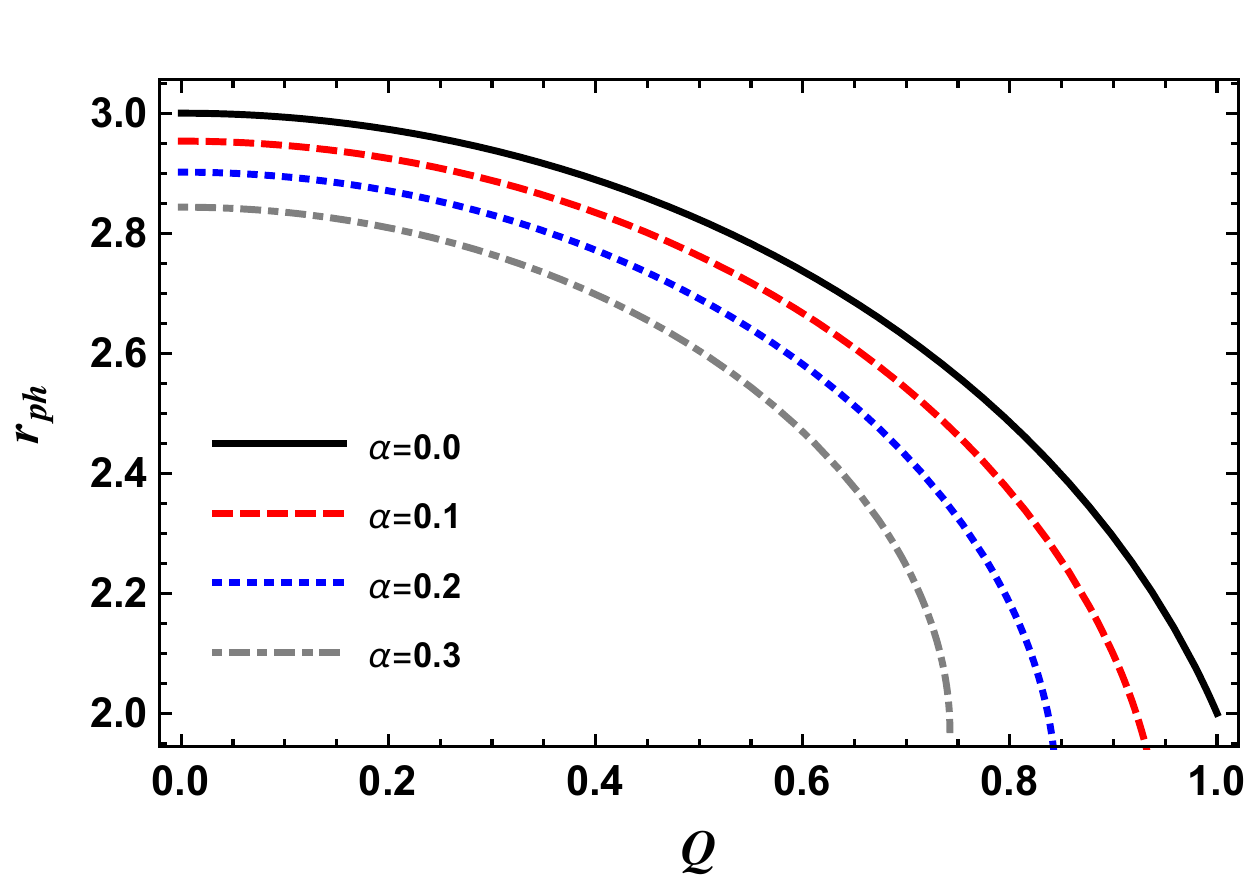}
  \end{center}
\caption{Plots showing the variation of radius of photon sphere with $\alpha$ the GB coupling constant and $Q$ charge for the non-rotating BH with $M=1$.}\label{photon plot}
\end{figure*}

In Fig. (\ref{photon plot}), we have plotted the radius of photon sphere with respect to $\alpha$, i.e., the GB coupling constant and $Q$ charge  using eq. (\ref{photon equation}). It can be seen that the outer horizon shifting towards the smaller value of $r$ for both the parameters, i.e., the GB coupling constant and BH charge which implies that the outer horizon shifting towards the central singularity. 

\subsection{Rotating black hole}
\label{rotating}
To obtain the rotating counterpart of charged BH in 4D EGB gravity we start with the metric (\ref{GBme}) to apply the NJA, using the approach followed in \cite{{Newman:1965tw,Newman:1965my}}. To begin with, let us introduce the metric (\ref{GBme}) in Eddington-Finkelstein coordinates ($u$, $r$, $\theta$, $\phi$) with

\begin{equation}\label{tr}
 du=dt-\frac{dr}{f(r)}.
\end{equation}

\begin{figure*}
 \begin{center}
   \includegraphics[scale=0.7]{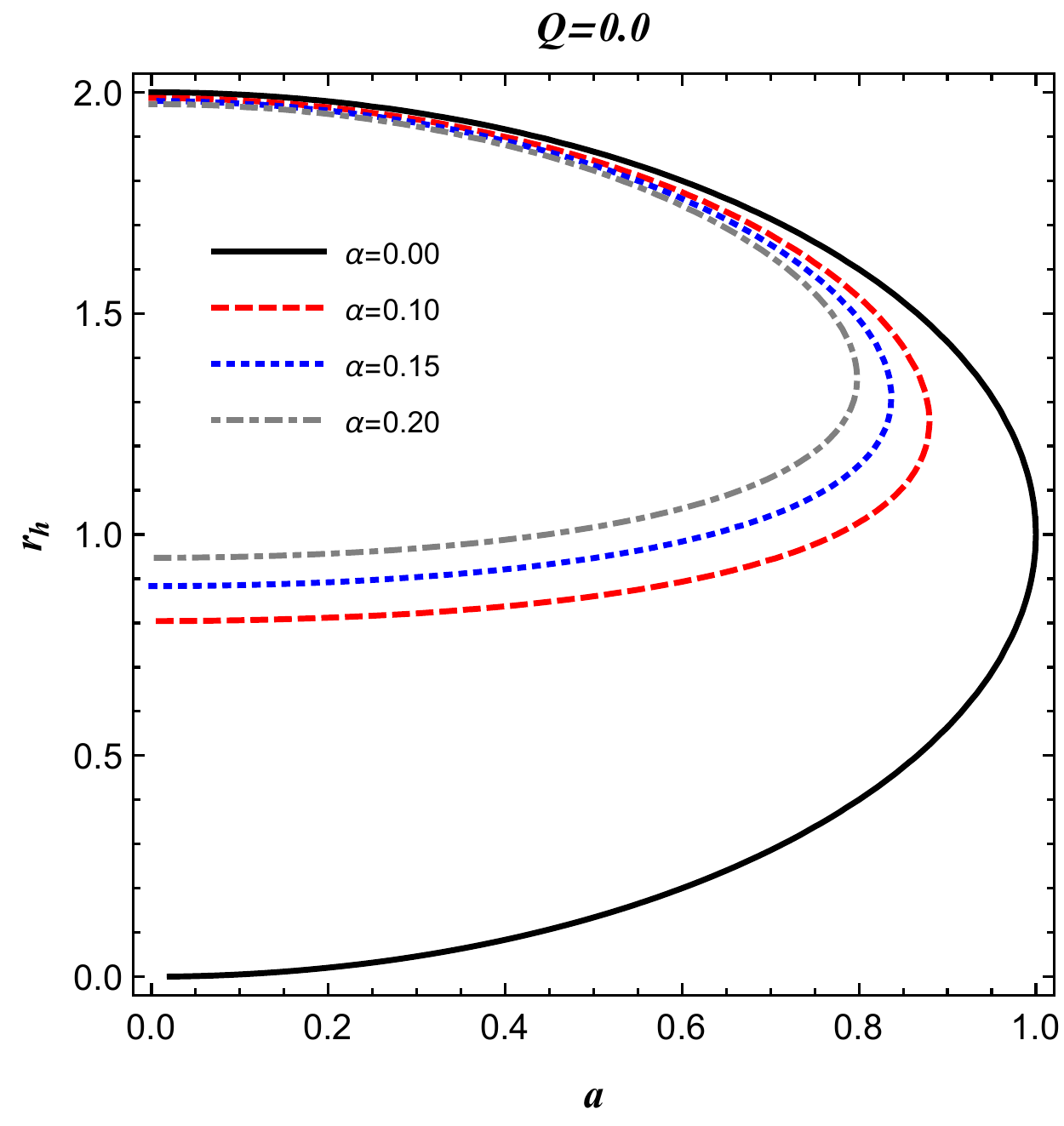}
   \includegraphics[scale=0.7]{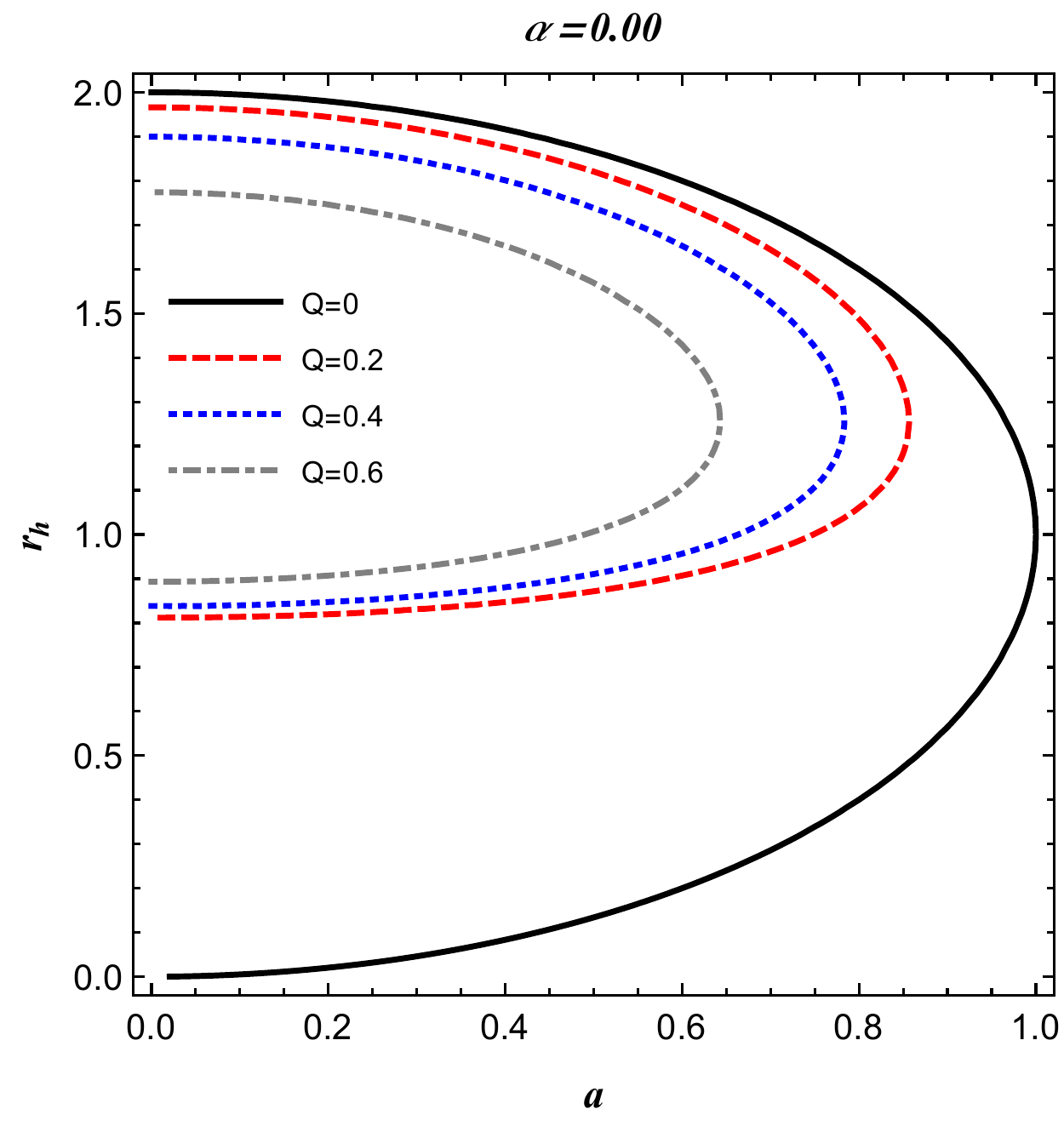}
   
   \includegraphics[scale=0.7]{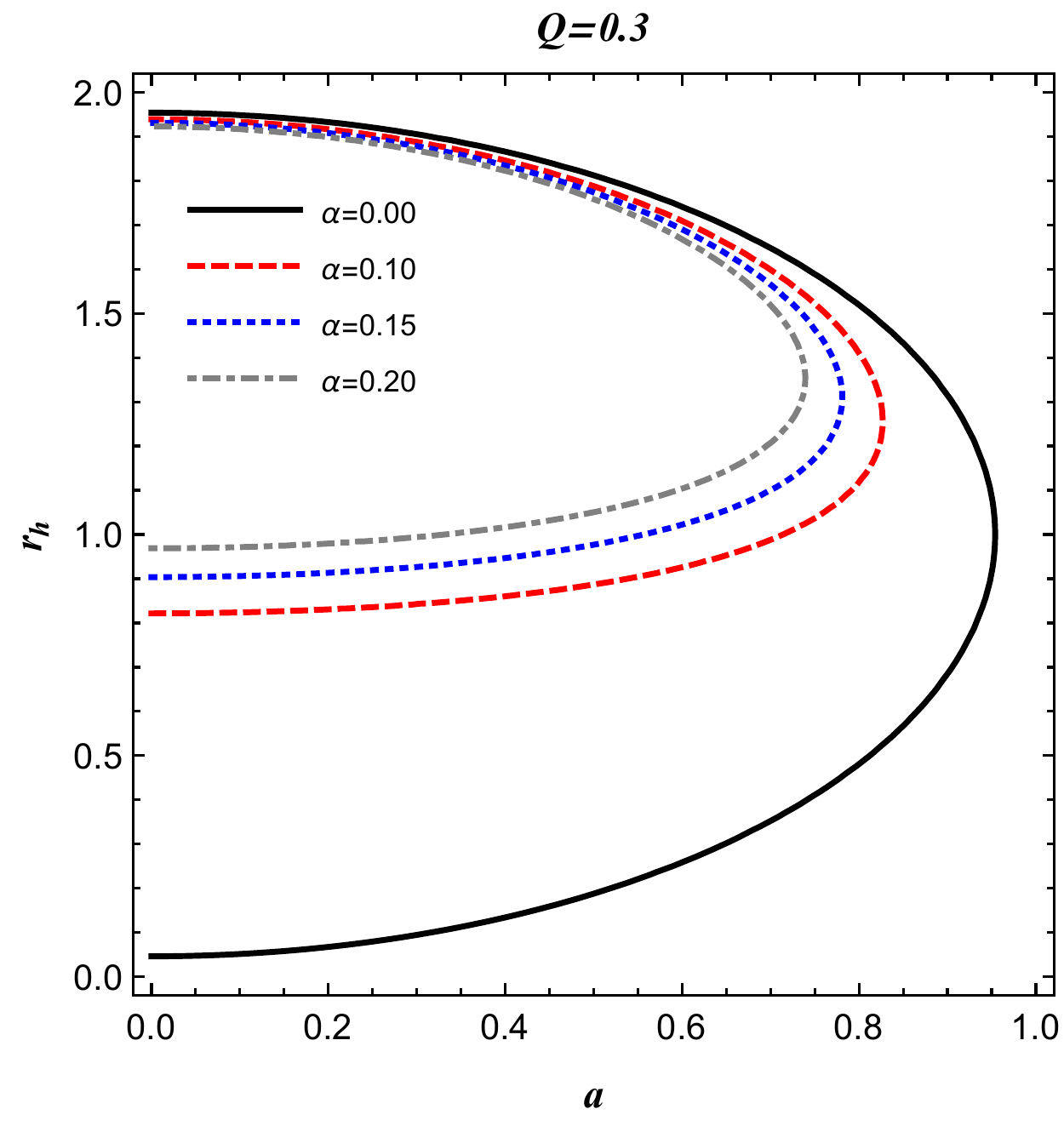}
   \includegraphics[scale=0.7]{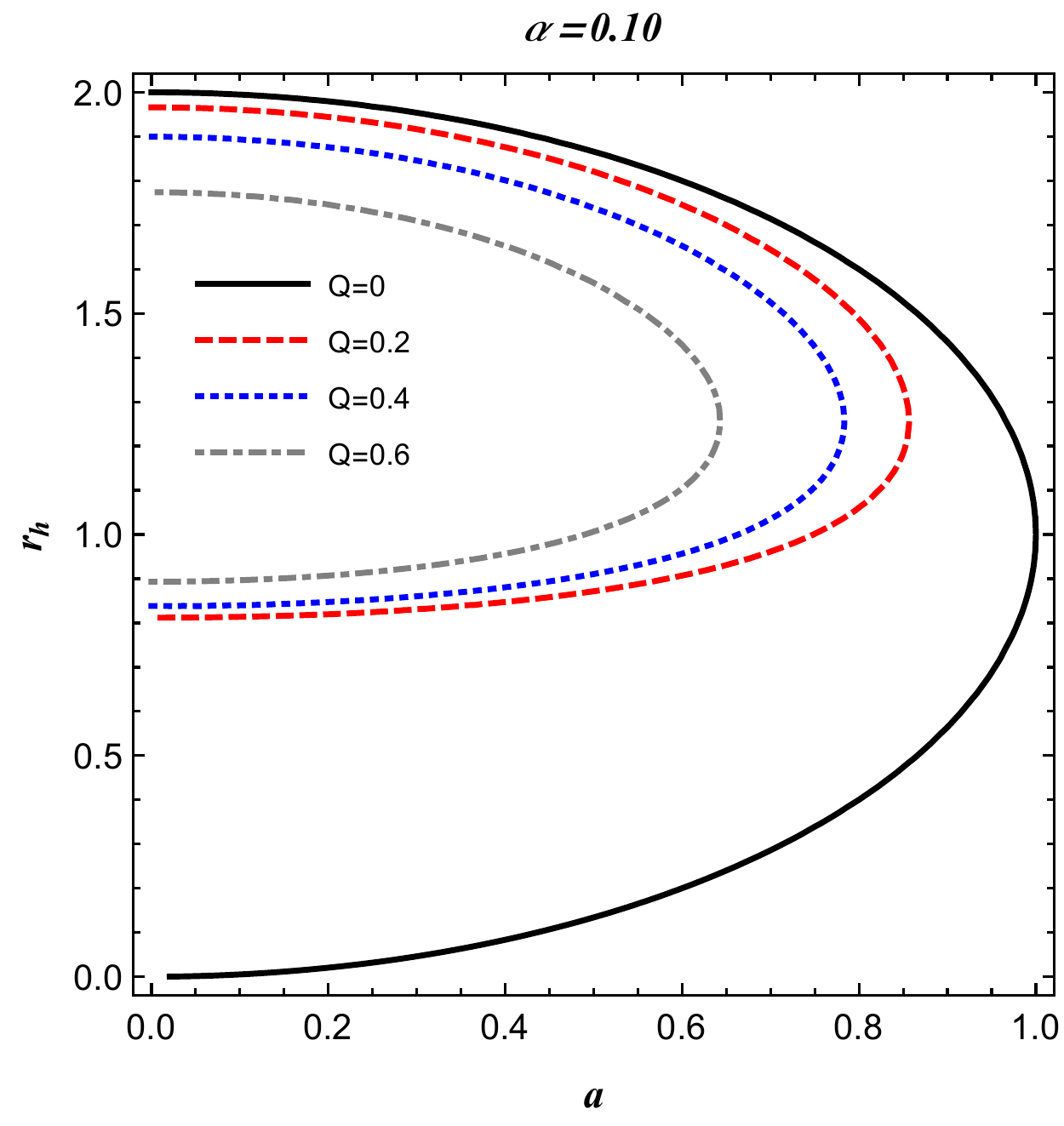}
  \end{center}
\caption{Variation of horizon(event and Cauchy) radius with spin parameter $a$ for different values of GB coupling constant $\alpha$ and charge $Q$ with $M=1$.}\label{hor}
\end{figure*}

Following the transformation (\ref{tr}), the metric of the non-rotating charged GB BH becomes
\begin{equation}
 ds^2=f(r)du^2+2dudr-r^2d\theta^2-r^2\sin^2\theta d\phi^2.\label{sur}
\end{equation}
Further, the metric can be expressed in terms of null tetrad as
\begin{equation}
 g^{ab}=l^am^b+l^bn^a-m^a\bar{m}^b-m^b\bar{m}^a.
\end{equation}
Note $l^a$ and $n^a$ are real, $m^a$, $\bar{m^a}$ are mutual complex conjugate. Now, following \cite{Newman:1965tw,Newman:1965my}, the null tetrad of the metric is of the form 
\begin{eqnarray}
 l^a&=&\delta^a_r,\\
 n^a&=&\delta^a_\mu-\frac{f(r)}{2}\delta^a_r,\\
 m^a&=&\frac{1}{\sqrt{2}r}\left(\delta^a_\theta+\frac{i}{\sin\theta}\delta^a_\phi\right).
\end{eqnarray}
The null tetrad obeys null, orthogonal, and metric
conditions as follows
\begin{eqnarray}
 l^al_a=n^an_a=m^am_a=\bar{m}^a\bar{m}_a=0,\\
 l^am_a=l^a\bar{m}_a=n^am_a=n^a\bar{m}_a=0,\\
 l^an_a=-m^a\bar{m}_a=1.
\end{eqnarray}
Now, following NJA, performing the complex coordinate transformations in ($u$, $r$)-plane using the relation
\begin{eqnarray}
 && u'\rightarrow u-ia\cos\theta,\nonumber\\
 && r'\rightarrow r+ia\cos\theta,\label{rp}
\end{eqnarray}
with $a$ as the spin parameter of the BH. Next, complexification of the radial coordinate results in the change in the metric functions of (\ref{sur}) to new undetermined ones. At the same time \cite{Mustafa2014,Mustafa2014a}
\begin{eqnarray}
 &&f(r)\rightarrow F(r, a, \theta),\\
 &&r^2\rightarrow H(r, a, \theta).
\end{eqnarray}
Following this transformation, the null tetrad with the spin parameter $a$ will becomes
\begin{eqnarray}\label{nt}
 l^a&=&\delta^a_r,\\
 n^a&=&\delta^a_\mu-\frac{F}{2}\delta^a_r,\\
 m^a&=&\frac{1}{\sqrt{2H}}\left((\delta^a_\mu-\delta^a_r)ia\sin\theta+\delta^a_\theta
 +\frac{i}{\sin\theta}\delta^a_\phi\right).
\end{eqnarray}
Thus, applying the new null tetrad (\ref{nt}) the charged rotating metric for 4D-EGB BH in the Eddington-Finkelstein coordinates obtained is given by
\begin{eqnarray}\label{mt}
 ds^2&=& Fdu^2+2du dr+2a\sin^2\theta(1-F)du d\phi \nonumber \\ && -2a\sin^2\theta dr d\phi -H d\theta^2 \nonumber\\
 && -\sin^2\theta\left(H+a^2\sin^2\theta(1-F)\right)d\phi^2.
\end{eqnarray}
Next step is to change the metric (\ref{mt}) into the Boyer-Lindquist coordinates. We obtain the rotating EGB BH by introducing  a global coordinate transformation\cite{Mustafa2014}:
\begin{eqnarray}
 du&=&dt+\nu(r)dr,\\
 d\phi&=&d\phi'+\chi(r)dr,
\end{eqnarray}
with \cite{Newman:1965tw}
\begin{eqnarray}
 \nu(r)&=&-\frac{a^2+r^2}{a^2+r^2f(r)},\\
 \chi(r)&=&-\frac{a}{a^2+r^2f(r)}.
\end{eqnarray}
By choosing
\begin{eqnarray}
 F&=&\frac{(r^2f(r)+a^2\cos^2\theta)}{H},\;\;\;
 H=r^2+a^2\cos^2\theta.
\end{eqnarray}
Hence, the charged rotating 4D-EGB BH metric reads as
\begin{eqnarray}
 ds^2 &=& -\frac{\Delta}{\rho^2}(dt-a\sin^2\theta d\phi)^2+\frac{\rho^2}{\Delta}dr^2+\rho^2d\theta^2
  \nonumber \\ && +\frac{\sin^2\theta}{\rho^2}\left(a dt-(r^2+a^2)d\phi\right)^2.\label{Romet}
\end{eqnarray}
For the metric (\ref{Romet}), the metric functions are
\begin{eqnarray}
 \rho^2&=&r^2+a^2\cos^2\theta,\\
 \Delta&=&r^2+a^2+\frac{r^2}{2\alpha}\left(1-\sqrt{1+4\alpha\Big(\frac{2 M}{r^3}-\frac{Q^2}{r^4}\Big)}\right).\label{DDm}
\end{eqnarray}
Note that, we have changed the sign of the metric according to the convention. In Eq. (\ref{DDm}), there are two branches of solutions for $\alpha >0$ and $\alpha <0$. Here we focus on the
case $\alpha > 0$ \cite{Glavan_2020}.

\begin{figure*}
 \begin{center}
   \includegraphics[scale=0.7]{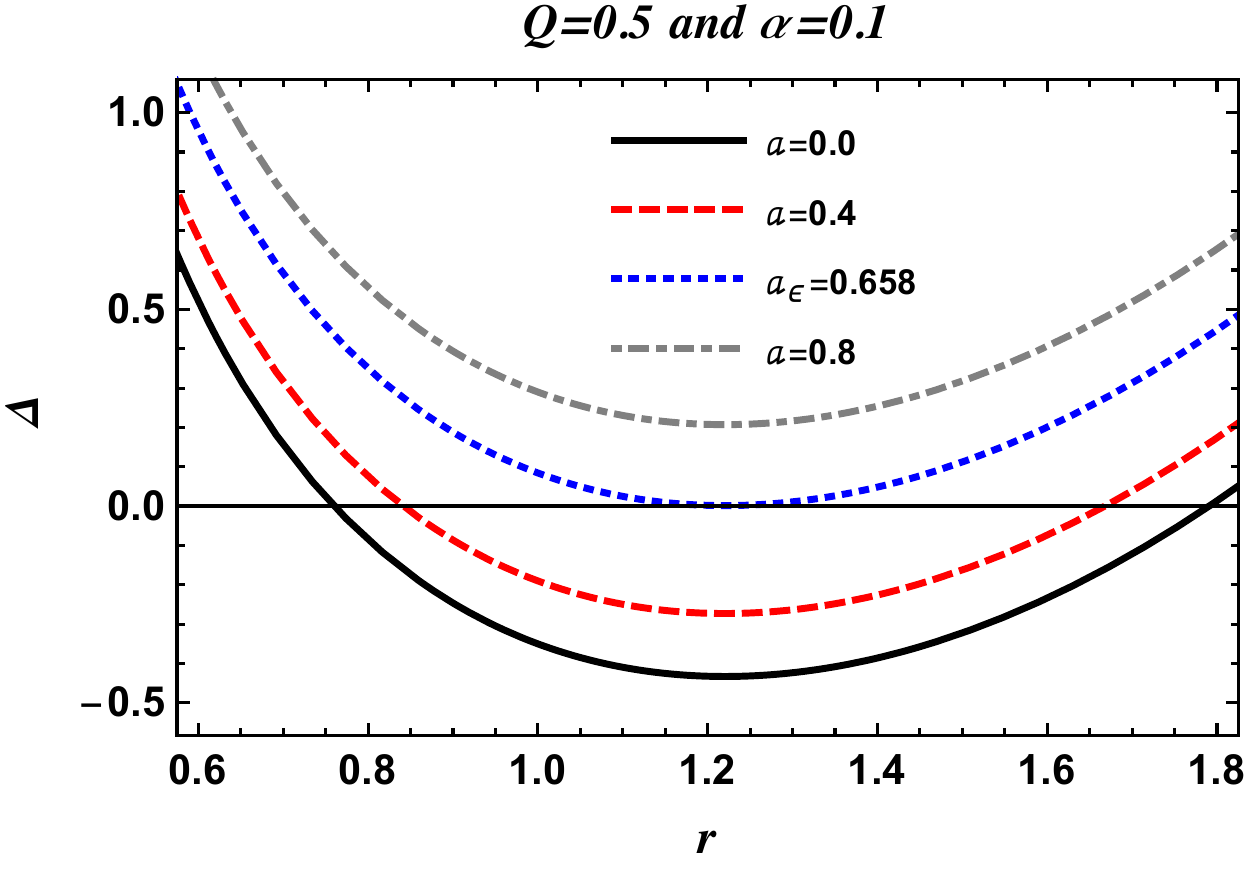}
   \includegraphics[scale=0.7]{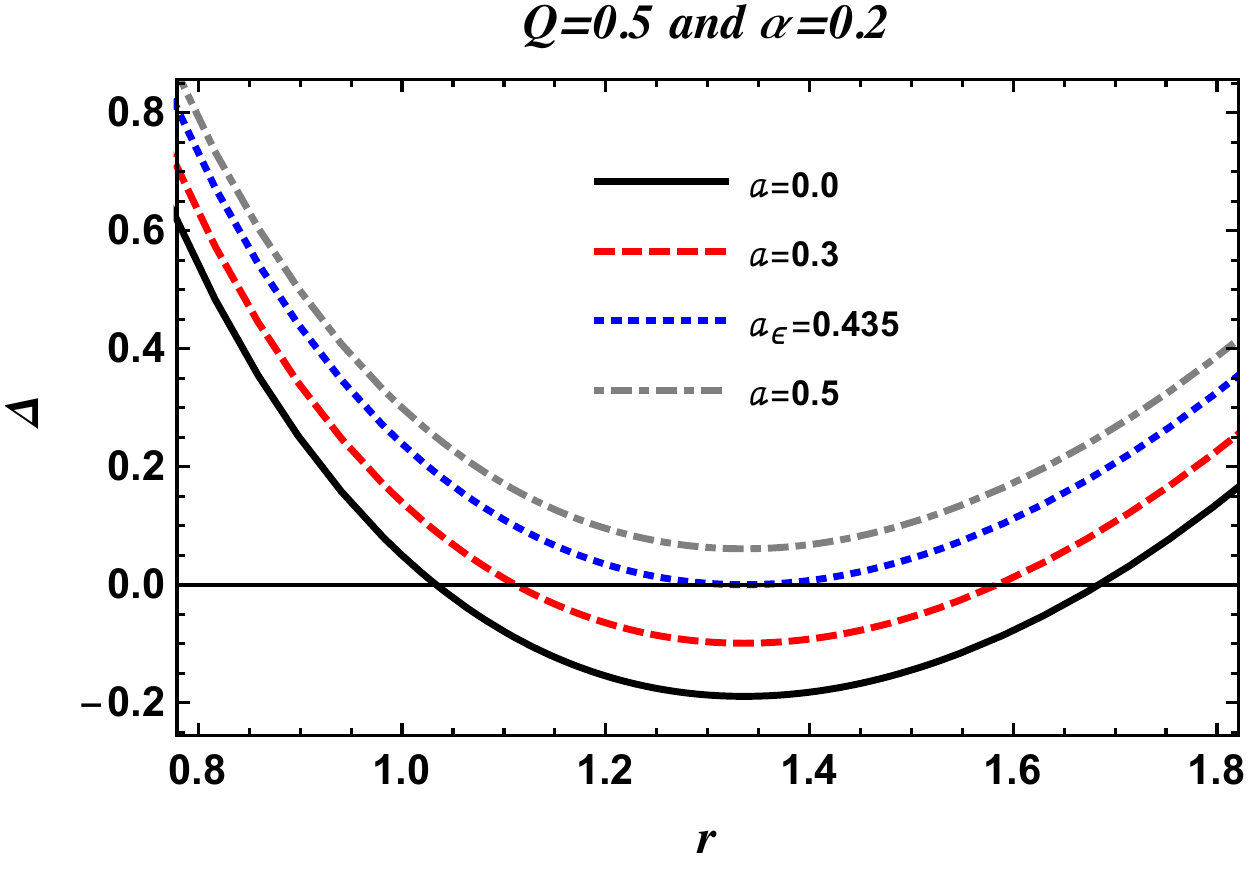}
   
   \includegraphics[scale=0.7]{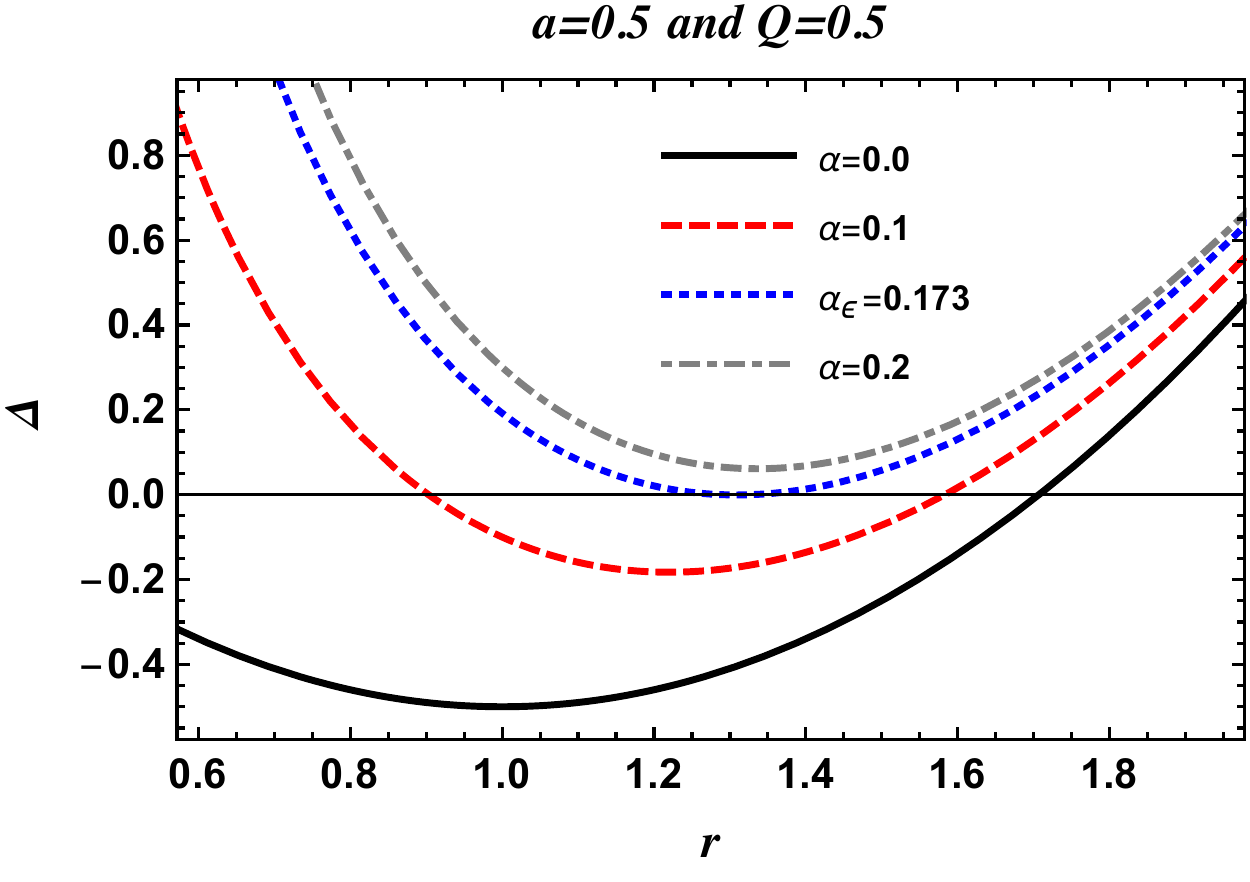}
   \includegraphics[scale=0.7]{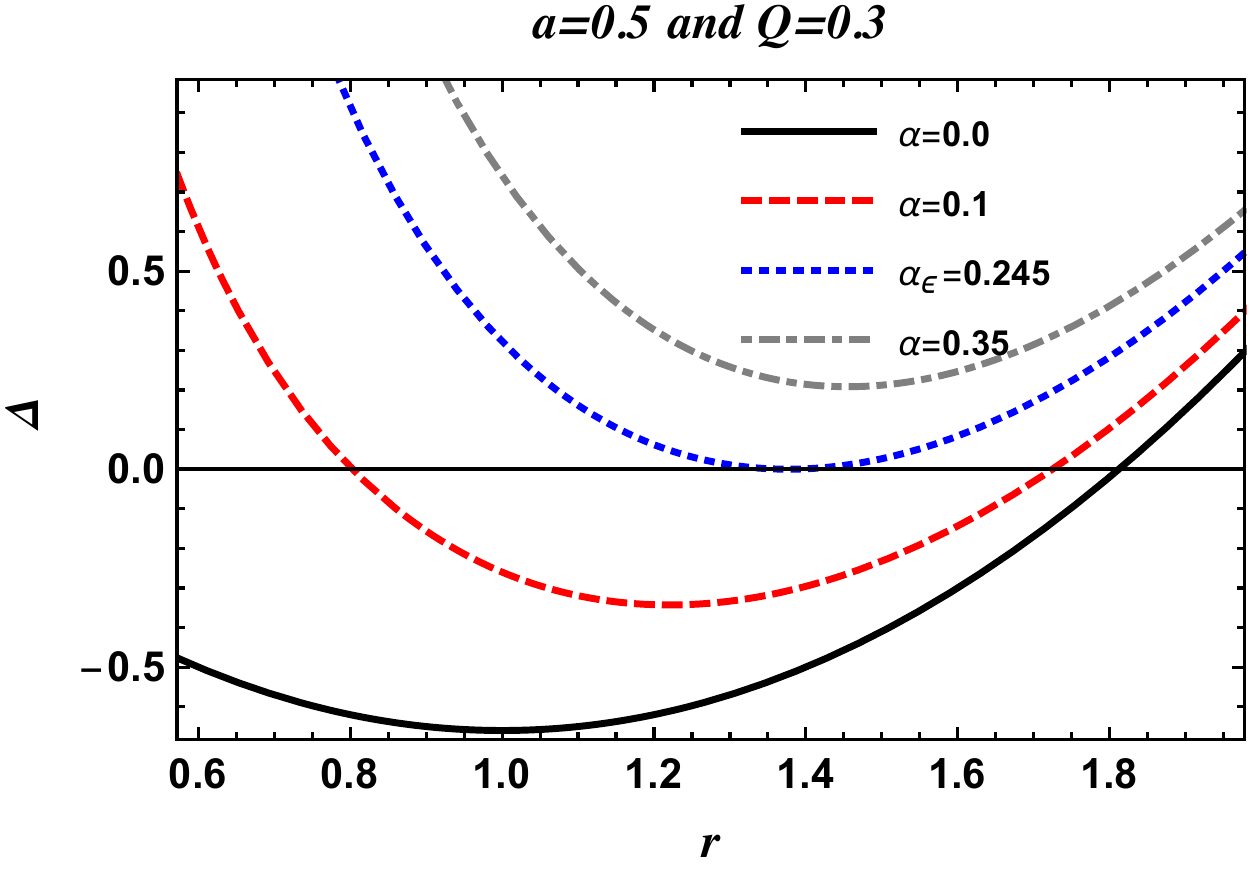}
   
      \includegraphics[scale=0.7]{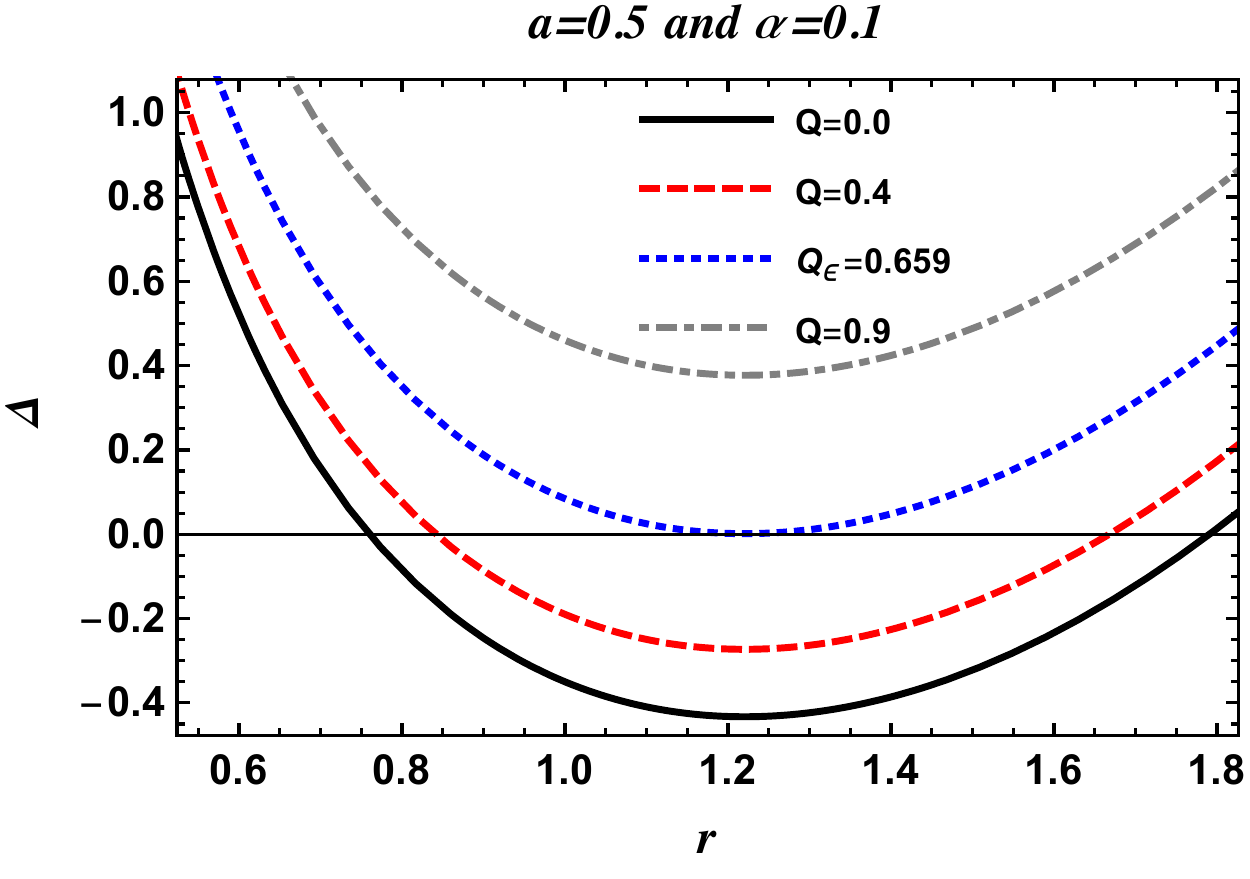}
   \includegraphics[scale=0.7]{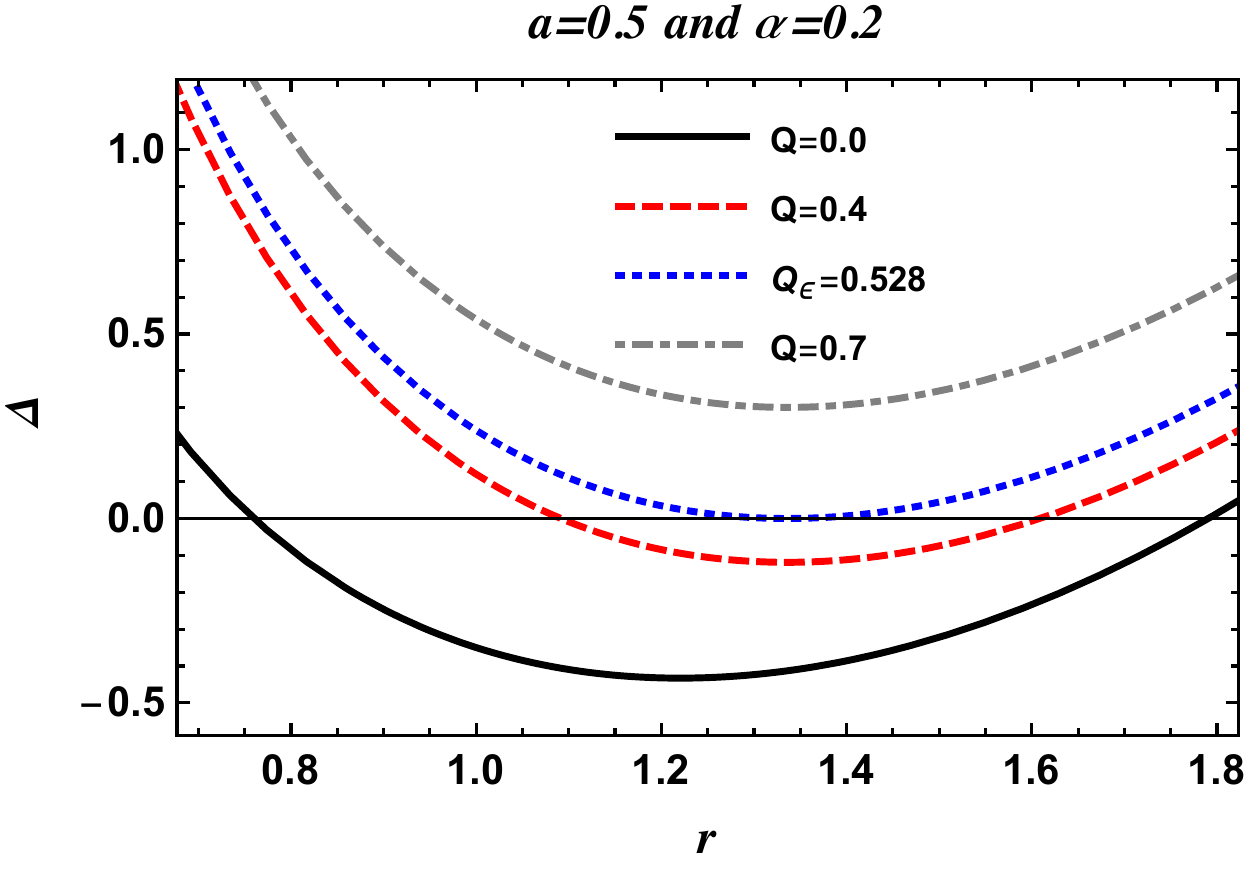}
  \end{center}
\caption{: (Upper panel) Plot showing the horizons for various values of spin parameter $a$ for fixed values of $\alpha$ and $Q$. (Middle Panel) Plot showing the horizons for various values of GB parameter $\alpha$ for fixed values of $a$ and $Q$.  (Lower
panel) Plot showing the horizons for different values of $Q$ for fixed  $\alpha$ and $a$ with $M=1$. The blue dashed lines correspond to the extremal BH and $a_{\epsilon}$, $\alpha_{\epsilon}$ and $Q_{\epsilon}$ are the corresponding extremal values.}\label{delta}
\end{figure*}

\subsection{Horizon structure}
\label{horizon}
Now we wish to discuss the effect of spin parameter, GB parameter and BH charge on the structure of horizons of charged rotating BH in 4D-EGB gravity. The BH spacetime (\ref{Romet}) corresponds to the horizon $\Delta =0$. We solve $\Delta =0$ for horizons numerically and the behaviour of horizon radius with respect to $a$ is shown in Fig. (\ref{hor}) for different values of parameters $\alpha$ and $Q$. It can be seen that the horizon
radius decreases 
with increasing $\alpha$ or $Q$, such that charged BHs of
4D EGB gravity always possess smaller horizon radius
as compared to the Kerr BH. 

Further, Fig. (\ref{hor}) and  Fig. (\ref{delta}) shows the existence of two roots, for a set of values of parameters $a$, $\alpha$ and $Q$, which corresponds to the Cauchy horizon (smaller root) and event horizon (larger root). From the horizon structure depicted
in Fig. (\ref{delta}), it turns out that, there exists critical extremal value of $a$, $\alpha$ and $Q$ for fixed $Q, \alpha$, $a, Q$ and $a, \alpha$ respectively, where $\Delta = 0$ has
a double root. The horizon for charged rotating BH in GB parameter exists at $a < a_{\epsilon}, \alpha < \alpha_{\epsilon}$ and $Q < Q_{\epsilon}$ while
for $a > a_{\epsilon}, \alpha > \alpha_{\epsilon}$ and $Q > Q_{\epsilon}$ there is a naked singularity. It can be also be seen that $a_{\epsilon}$ and $Q$ increases with increasing the value of GB parameter whereas $\alpha_{\epsilon}$ increases with decreasing the value of $Q$ with fixed $Q$, $a$ respectively. For fixed values of $Q$, $a$ and $\alpha$,  event horizon radius is decreasing and cauchy horizon radius is increasing with the increase in $a, \alpha$ and $Q$. Thus the charged rotating BH have smaller event horizon radii than the Reissner Nordst$\Ddot{o}$rm with and without GB parameter. At
the extremal value for charge, rotation parameter and GB parameter, the two horizons coincide and the radius
of this extremal horizon increases with the increase in GB
parameter and rotation parameter and decrease in charge parameter.

\begin{figure*}
 \begin{center}
   \includegraphics[scale=0.7]{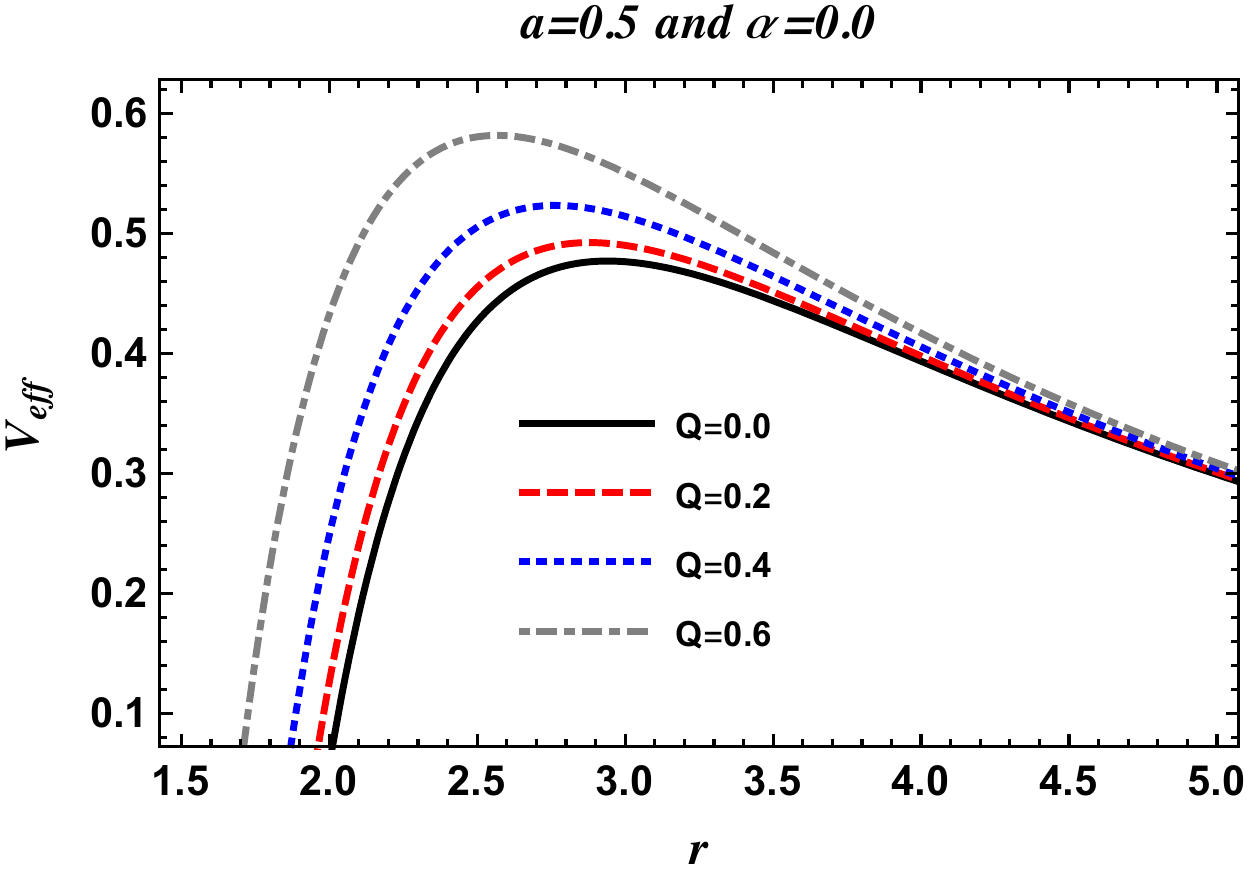}
   \includegraphics[scale=0.7]{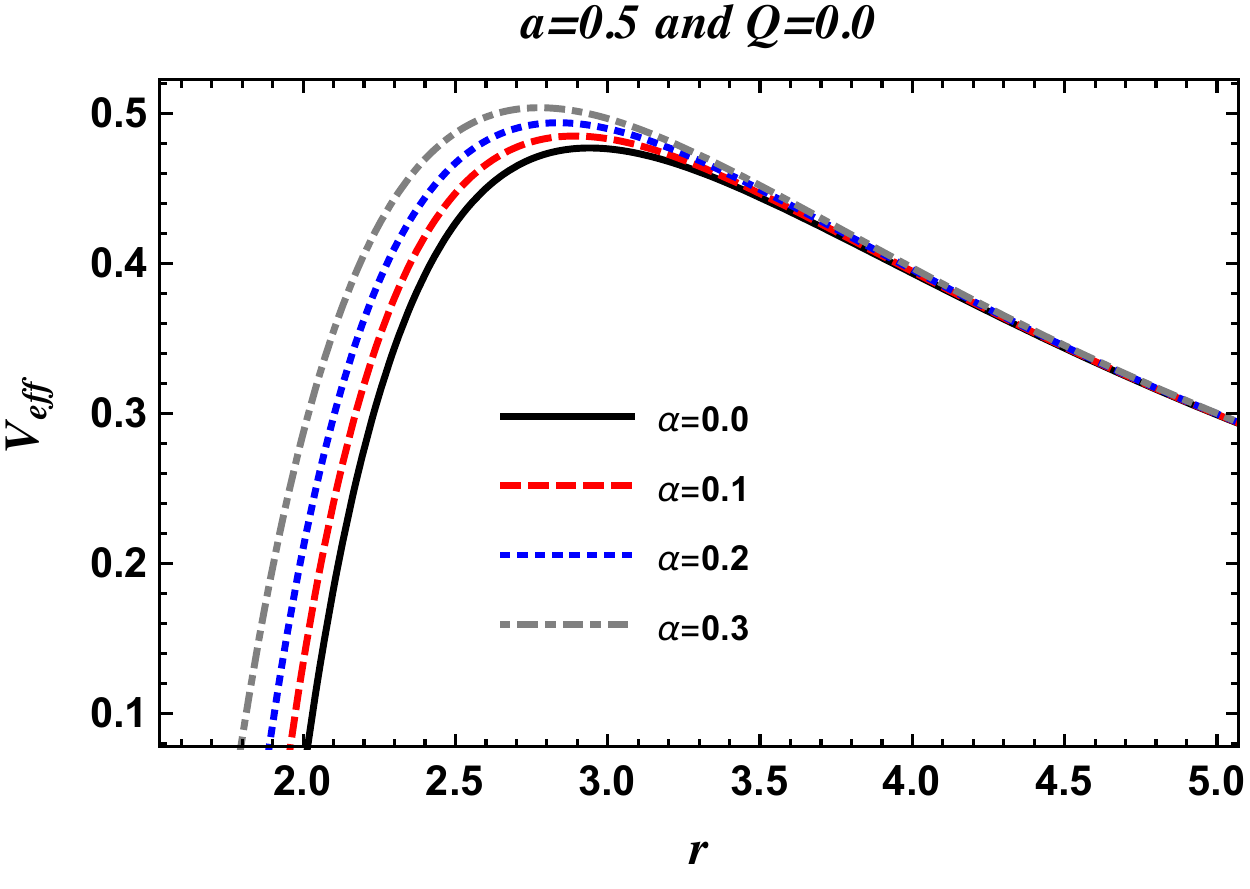}
   
   \includegraphics[scale=0.7]{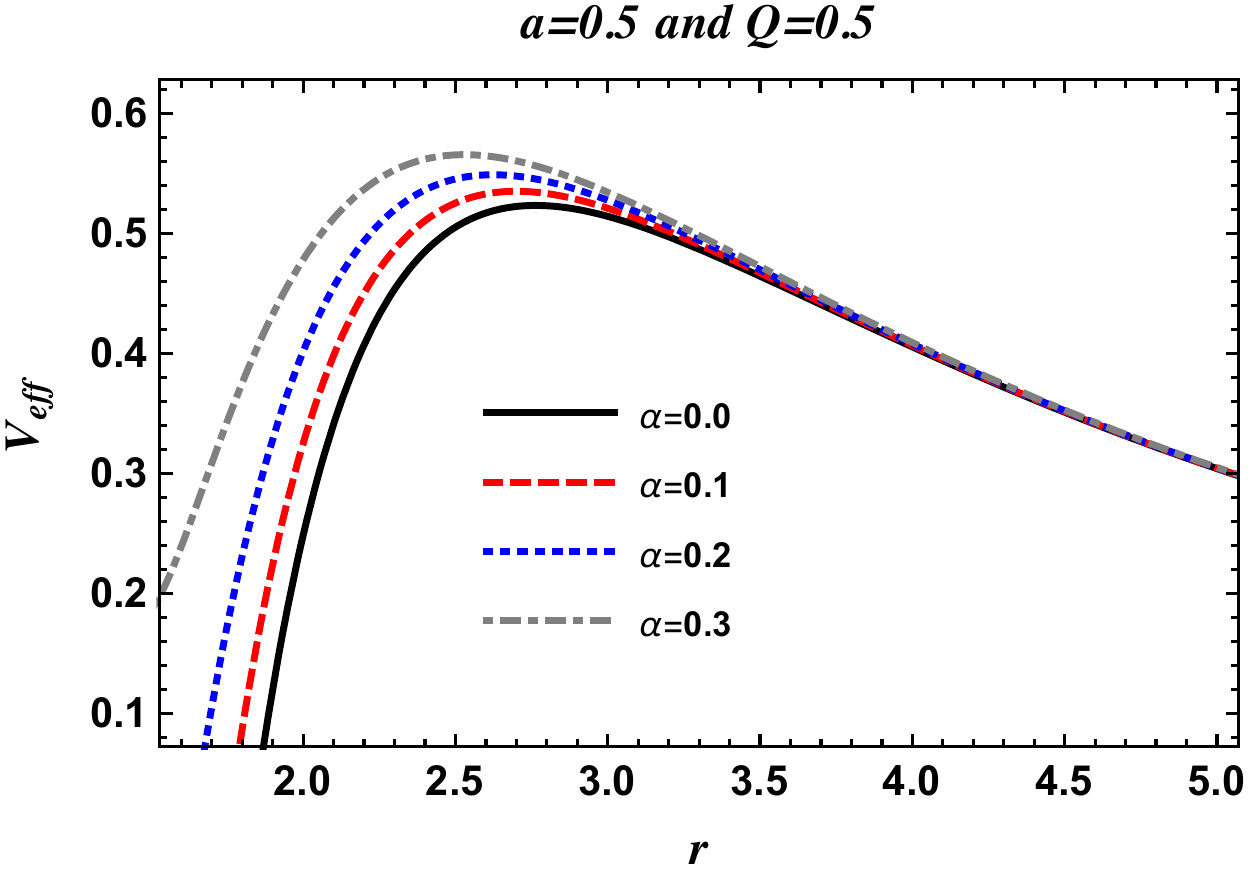}
   \includegraphics[scale=0.7]{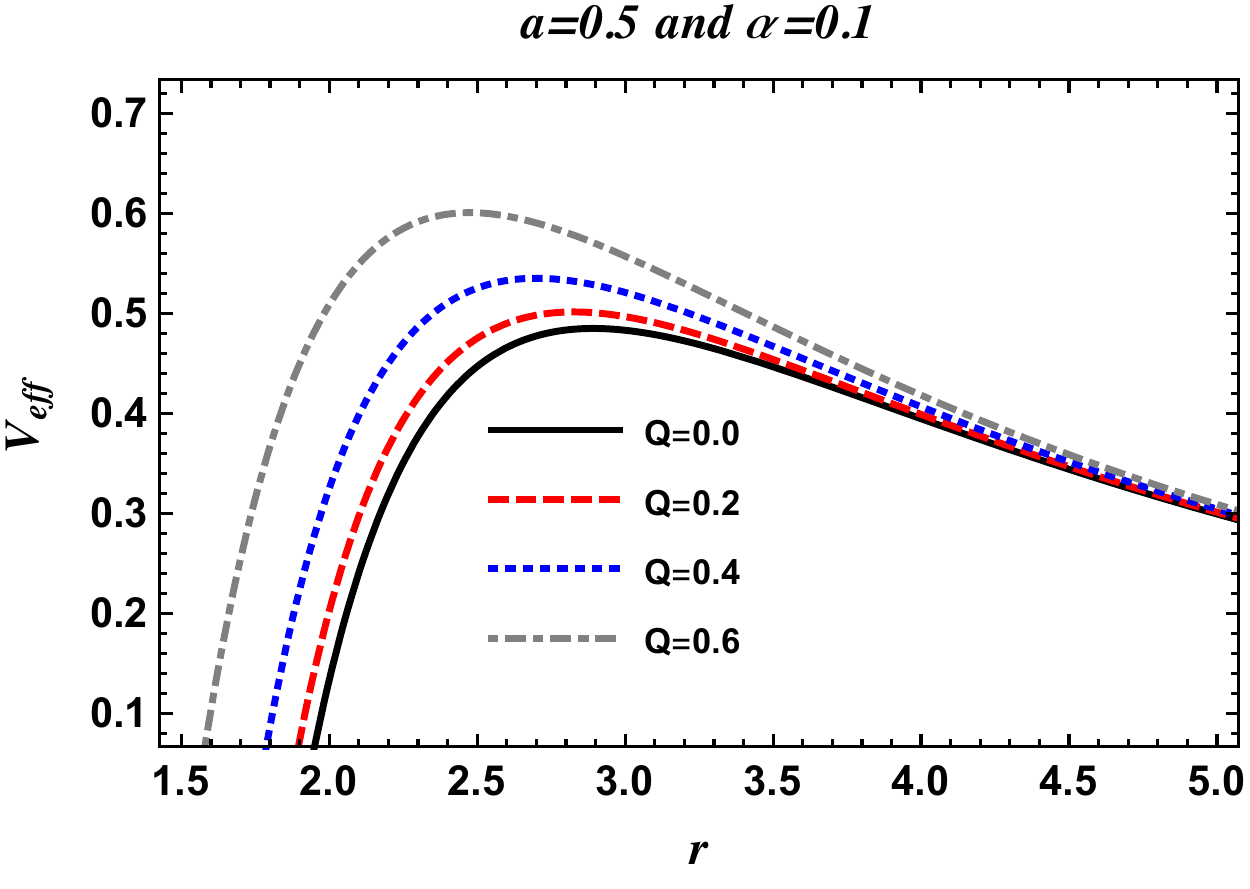}
   
   \includegraphics[scale=0.7]{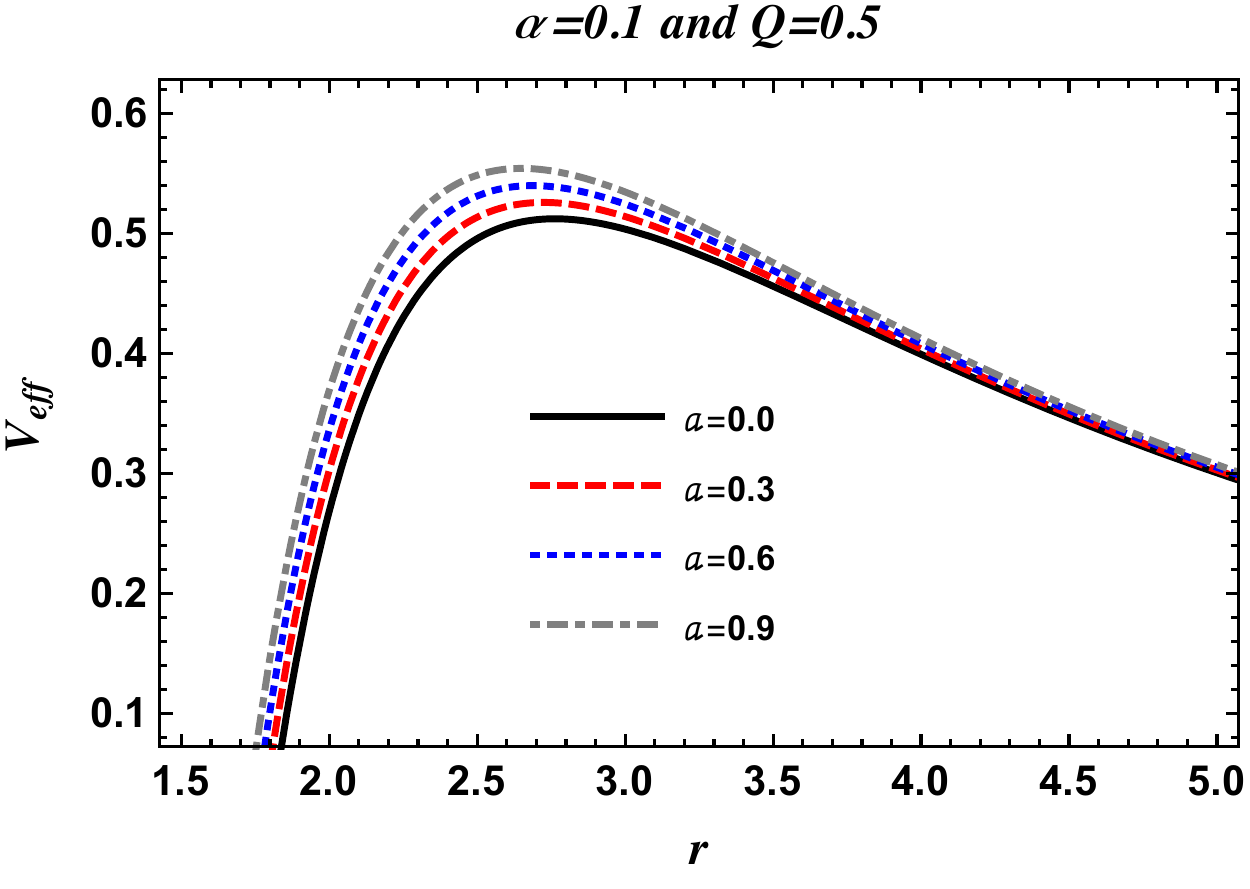}
   \includegraphics[scale=0.7]{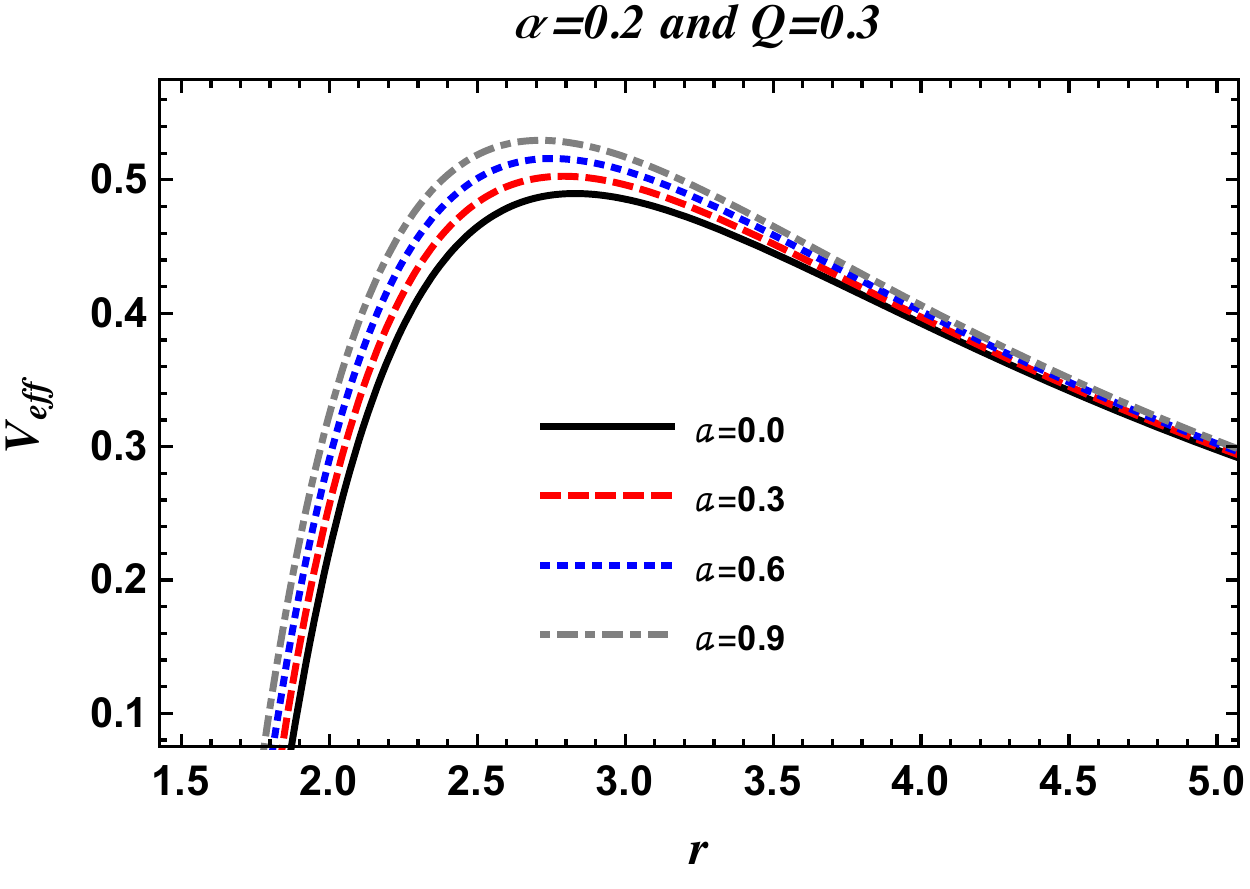}
  \end{center}
\caption{Plots showing the radial dependence of the effective potential for different values of $a$, $Q$ and $\alpha$ with $M=1$. }\label{vefflast}
\end{figure*}

\subsection{Null Geodesics}
\label{geodesics}
 In order to obtain the apparent shape of the shadow of the charged rotating 4D-EGB BH and discuss its properties we begin with the geodesic structure of photon in the background of charged rotating 4D-EGB BH. These geodesics can be determined by the Hamilton-Jacobi formulation \cite{PhysRev.174.1559} given by

\begin{figure*}
 \begin{center}
   \includegraphics[scale=0.7]{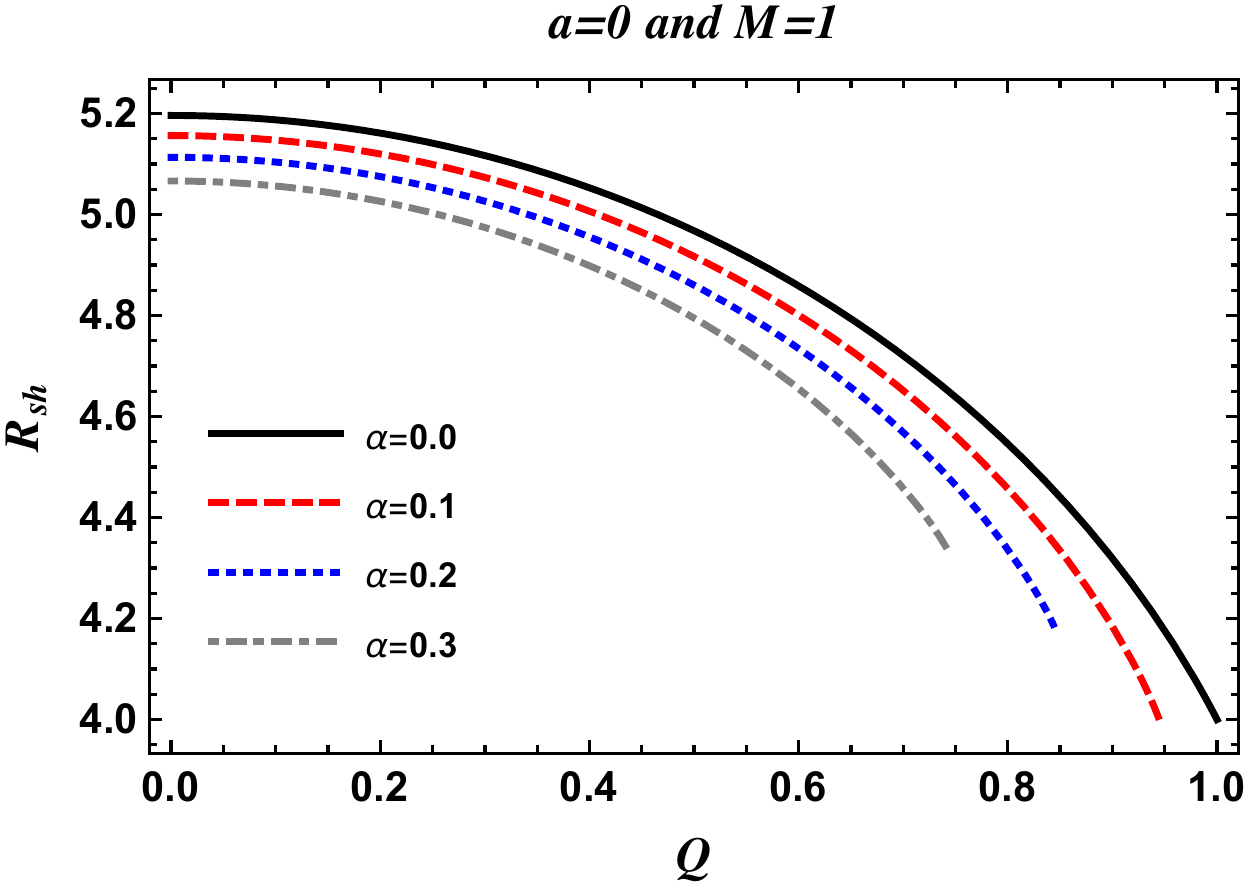}
   \includegraphics[scale=0.7]{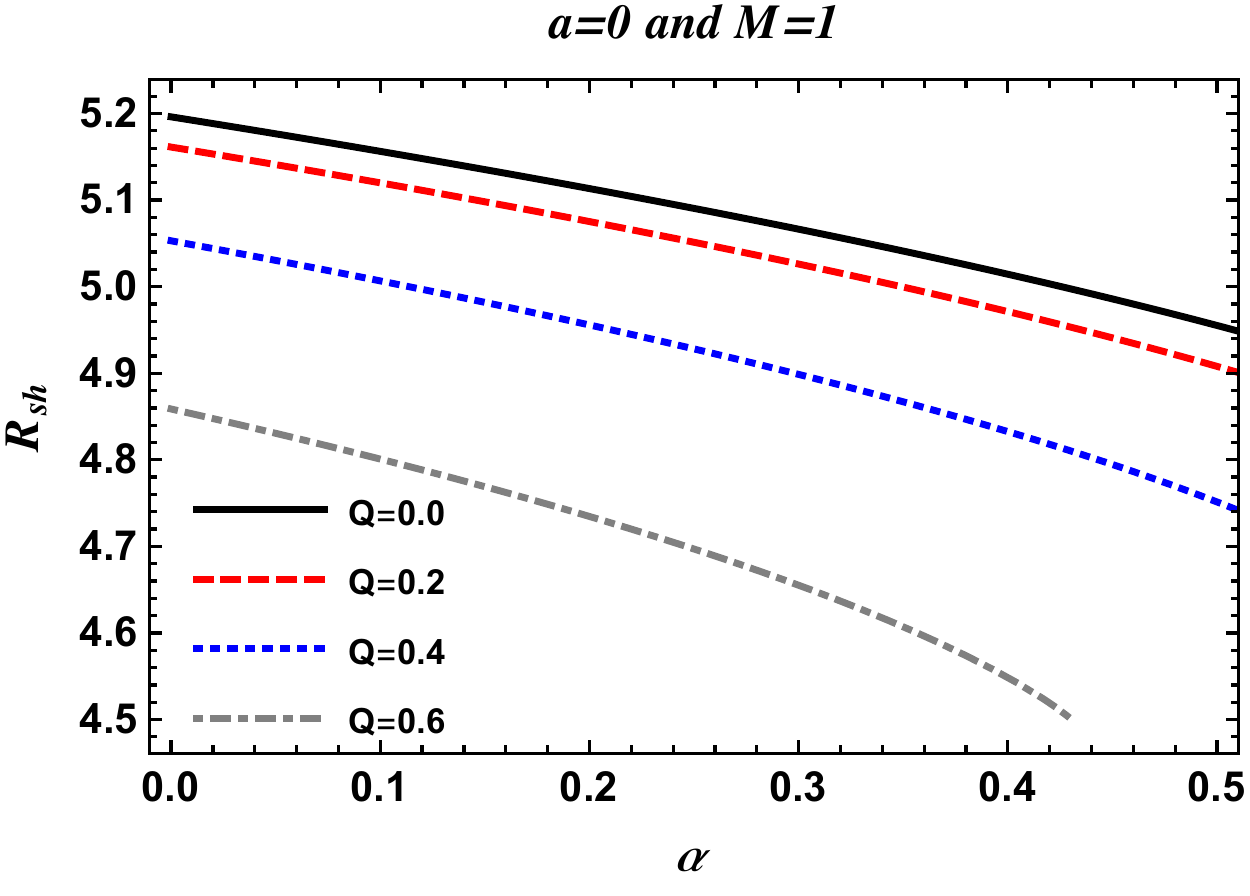}
  \end{center}
\caption{Plots showing the shape of the shadow of the black hole for the non-rotating case. }\label{shadownon plot}
\end{figure*}

\begin{equation}
 \frac{\partial S}{\partial\tau}=-\frac{1}{2}g^{\mu\nu}\frac{\partial S}{\partial x^{\mu}}\frac{\partial S}{\partial x^{\nu}},\label{sequation}
\end{equation}
\begin{eqnarray}
 -E&=&g_{t\mu}\dot{x}^{\mu},\\
l&=&g_{\phi\mu}\dot{x}^{\mu}.
\end{eqnarray}
where $E$ and $l$ are constant of motion corresponding to energy and momentum. $\tau$ is an affine parameter, $g^{\mu\nu}$ is the metric tensor. The action $S$ can be separated as
\begin{equation}
 S=\frac{1}{2}m^2\tau-Et+l\phi+S_{r}(r)+S_{\theta}(\theta),\label{jaction}
\end{equation}
where $m$ is the mass of the particle and $S_r(r)$, $S_{\theta}(\theta)$ are the function of $r$ and $\theta$. Using the
variable separable method and insert the Eq. (\ref{jaction}) into the Eq. (\ref{sequation}), we obtain the equations
of motion for photon $(m = 0)$

\begin{eqnarray}
 \mathcal{R}&=&\Big[(r^2+a^2)E-al\Big]^2-\Delta\Big[\mathcal{K}+(l-aE)^2\Big],\\
 \Theta&=&\mathcal{K}+\cos^2\theta\left(a^2 E^2-l^2\sin^{-2}\theta\right).
\end{eqnarray} 

The geodesic equations obtained are
\begin{eqnarray}
 \rho^{2}\frac{dt}{d\tau}&=&a(l-aE\sin^{2}\theta)
       \nonumber \\ &&+\frac{r^{2}+a^{2}}{\Delta}\Big(E(r^{2}+a^{2})-al\Big),\label{rhot}\\
 \rho^{2}\frac{dr}{d\tau}&=&\pm\sqrt{\mathcal{R}},\label{Rad}\\
 \rho^{2}\frac{d\theta}{d\tau}&=&\pm\sqrt{\Theta},\\
 \rho^{2}\frac{d\phi}{d\tau}
     &=&(l\csc^{2}\theta-aE)+\frac{a}{\Delta}\Big(E(r^{2}+a^{2})-al\Big).\label{rhophi}
\end{eqnarray}
For obtaining the boundary of the
BH shadow we study the radial equation. The radial equation of motion can be rewritten as 
\begin{equation}
 \left(\rho^{2}\frac{dr}{d\tau}\right)^2+V_{eff}=0.
\end{equation}
The effective potential in the equatorial plane reads as
\begin{eqnarray}\label{veff}
V_{eff}&=&\frac{\Delta ({\cal K} + (L - a E)^2)-((a^2 + r^2) E - a L)^2}{2r^4}.
\end{eqnarray}

In Fig. (\ref{vefflast}), we have shown the general behaviour of effective potential with respect to radius for different values of spin parameter, charge and GB parameter. It is seen that the effective potential exhibit a peak which correspond to unstable circular orbit. It is also seen that peak is increasing and shifting towards left with increase in the value of $a$, $Q$ and $\alpha$ which signifies the shifting of circular orbits to central object.

\section{Black hole shadow and observables}
\label{observa}
To determine the shape of the shadow of a charged rotating 4D-EGB BH, Let us define the following impact parameters $\xi = L/E$ and $\eta = \mathcal{K} /E^2$. Thus, $R((r)$ in terms of new parameters can be written as 

\begin{eqnarray} \label{r}
 \mathcal{R}&=&\Big[(r^2+a^2)-a\xi\Big]^2-\Delta\Big[\eta+(\xi-a)^2\Big].
 \end{eqnarray} 

Since, the orbits of the photons coming towards the BH follow three trajectories, i.e., falling into the BH, scattering away from the BH or making a circular orbit near the BH out of which the
unstable circular orbit near the BH is responsible for determining the shape of the shadow. The unstable circular photon orbits can be obtained by the following conditions

\begin{equation}
 V_{eff}=0,\quad \frac{\partial V_{eff}}{\partial r}=0, \quad \frac{\partial^2 V_{eff}}{\partial r^2}<0, \label{qqq}
\end{equation}or
\begin{equation}
R(r)=0=\frac{\partial{R(r)}}{\partial{r}}.
\end{equation}
The parameters $\xi$ and $\eta$ determine the shape of the shadow. Then, using  Eq. (\ref{r}) and (\ref{qqq}) we obtain $\xi$ and $\eta$ as

\begin{eqnarray}
 \xi&=&\frac{\left(a^2+r^2\right) \Delta'-4 \Delta r}{a \Delta'},\label{xx1}\\
 \eta&=&\frac{r^2 \left(16 \Delta  \left(a^2-\Delta\right)-r^2 \Delta'^2+8
   \Delta  r \Delta'\right)}{a^2 \Delta'^2},\label{xx2}
\end{eqnarray}
\begin{eqnarray}
 r+2\frac{\Delta}{\Delta'^2}(\Delta'-r\Delta'')>0,\label{ccdc}
\end{eqnarray}

\subsection{Nonrotating black hole shadows}
Using Synge method we can study shadow of black hole for the non-rotating case:

\begin{eqnarray}\label{shadow nonrotating1}
\sin^2 X_{s}=\frac{\gamma(r_{ph})^2}{\gamma(r_o)^2}.
\end{eqnarray}

where $X_{s}$ is the angular radius of the black hole shadow and $r_{o}$ is the observe position and $r_{ph}$ is the photon sphere.

For the large distance observation Eq. (\ref{shadow nonrotating1}) will be as 

\begin{eqnarray}\label{shadow nonrotating2}
\sin^2 X_{s}=\frac{r_{ph}^2}{f(r_{ph})}\frac{f(r_o)}{r^2_o},
\end{eqnarray}

And now one can find linear radius of BH shadow for large distance observation using Eq. (\ref{shadow nonrotating2}) as

\begin{eqnarray}\label{shadow nonrotating3}
R_{sh}=r_{o} \sin X_{s} = \frac{r_{ph}}{\sqrt{f(r_{ph})}}.
\end{eqnarray}

Finally Eq. (\ref{shadow nonrotating3}) can explain shadow of BH for the non-rotating case.

In Fig. (\ref{shadownon plot}), we have shown the graphical representation of shadow radius of charged 4D-EGB BH for different values of BH charge and GB parameter. It is worth to notice that the shadow decreases with the increase in value of both $Q$ and $\alpha$. It is evident that the shadows of the charged 4D-EGB BHs are smaller as compared to the Reissner Nordst$\Ddot{o}$rm BH and Schwarzschild  BH shadow as depicted by the black solid curves in Fig. (\ref{shadownon plot}). It is because the GB term acts as a repulsive gravitational charge, thus weakening the strength of the gravitational field, which allows the circular orbits to remain stable closer to the source.

\begin{figure*}
 \begin{center}
   \includegraphics[scale=0.45]{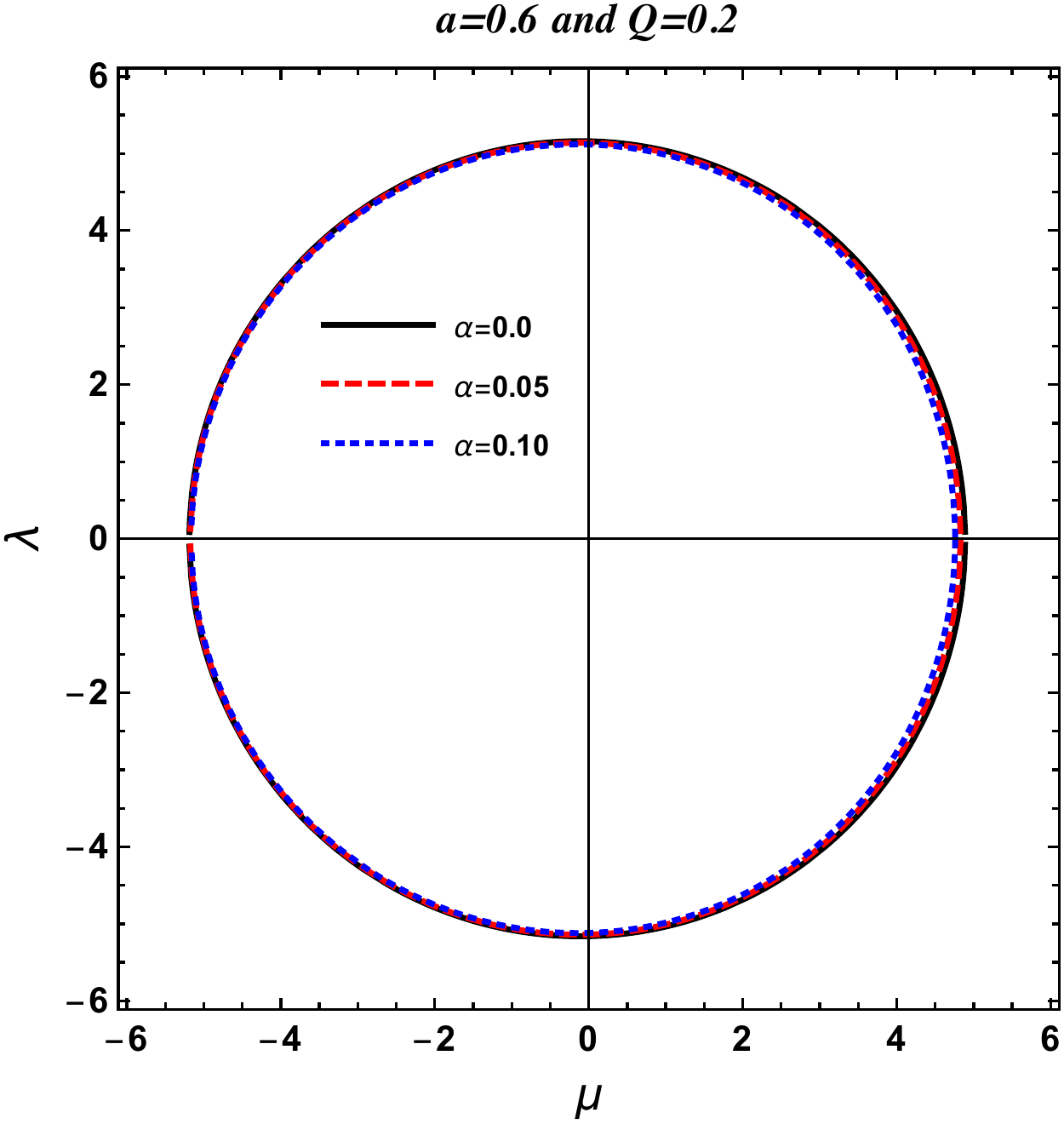}
   \includegraphics[scale=0.45]{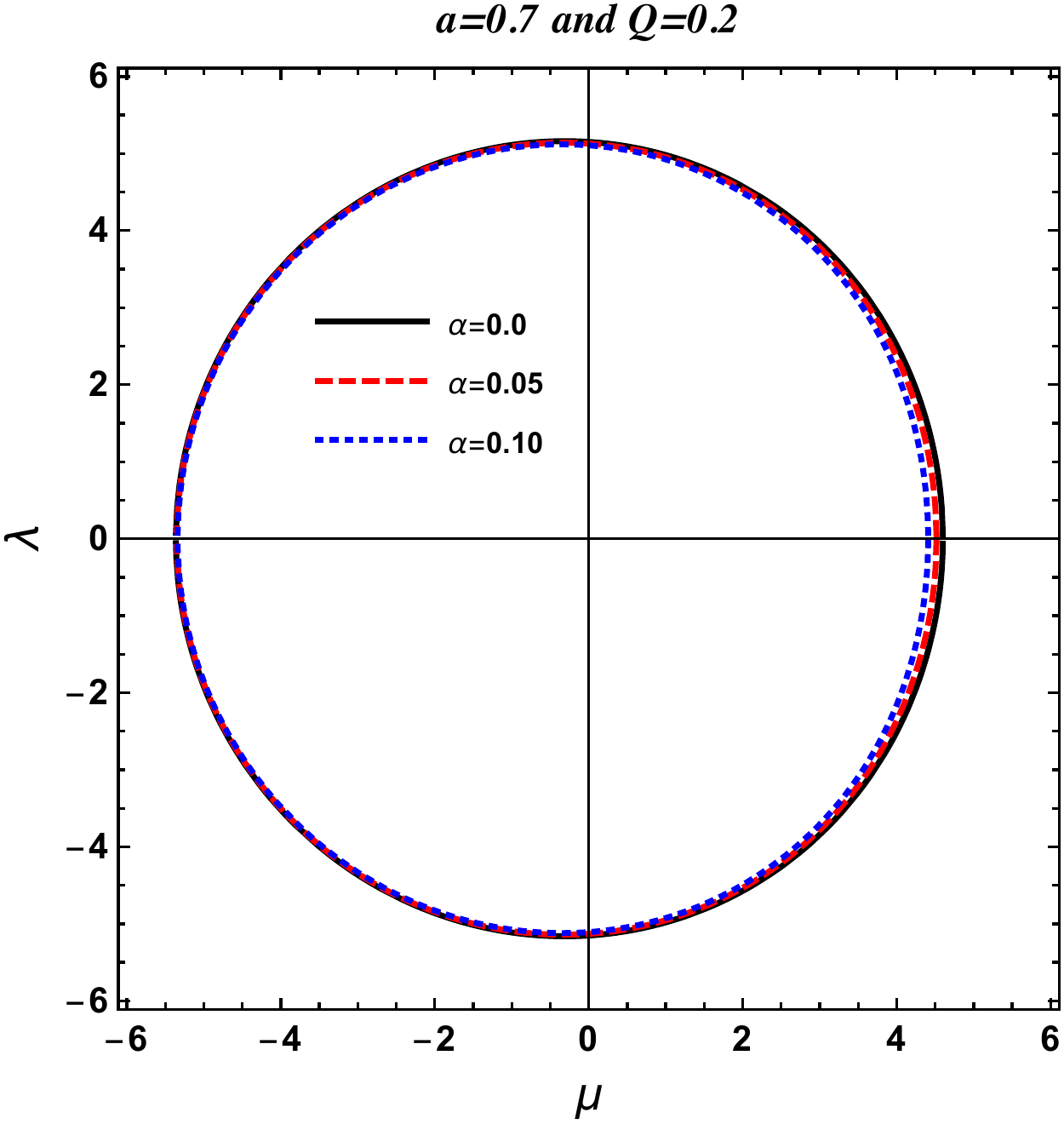}
   \includegraphics[scale=0.45]{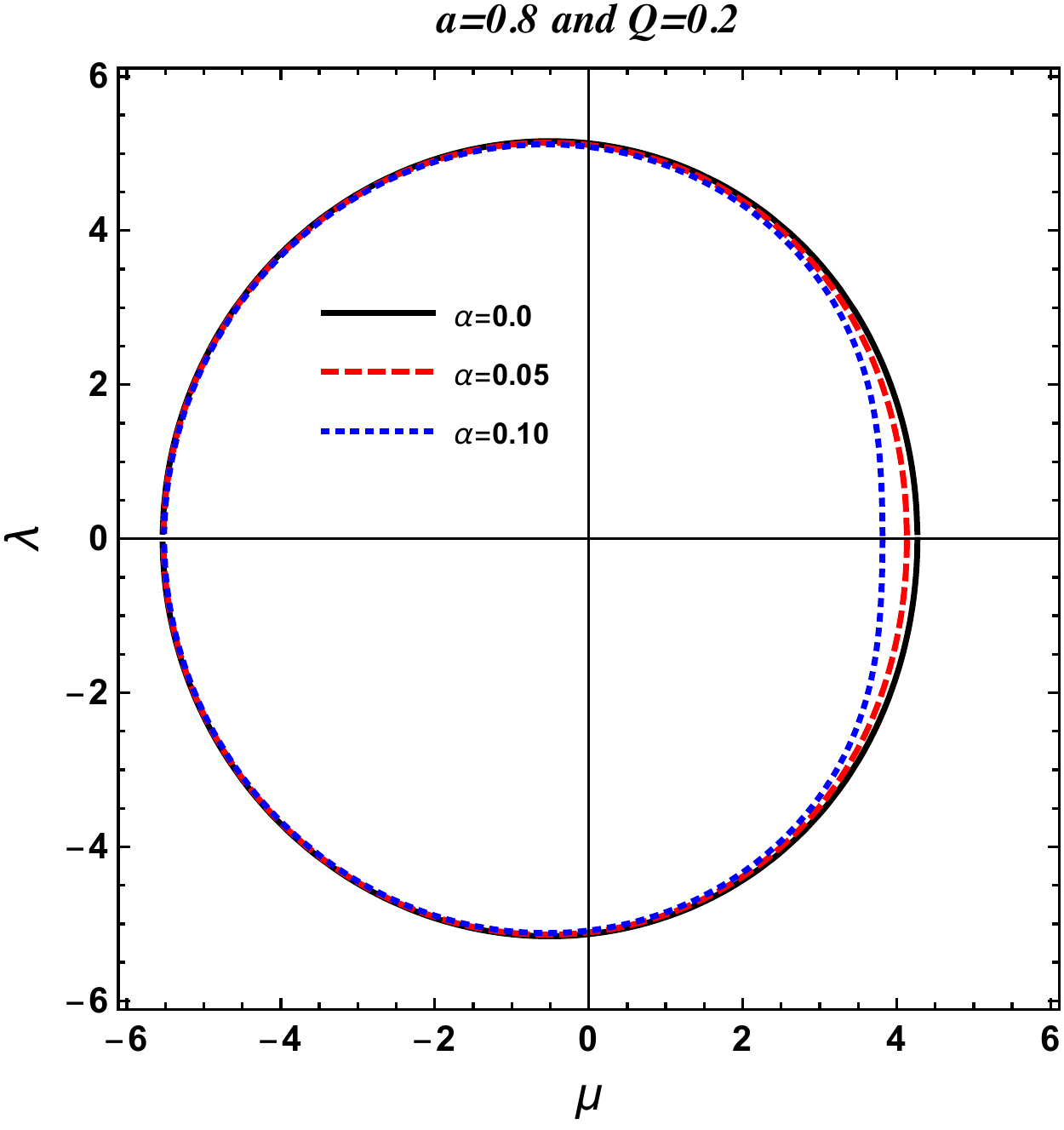}
   
   \includegraphics[scale=0.45]{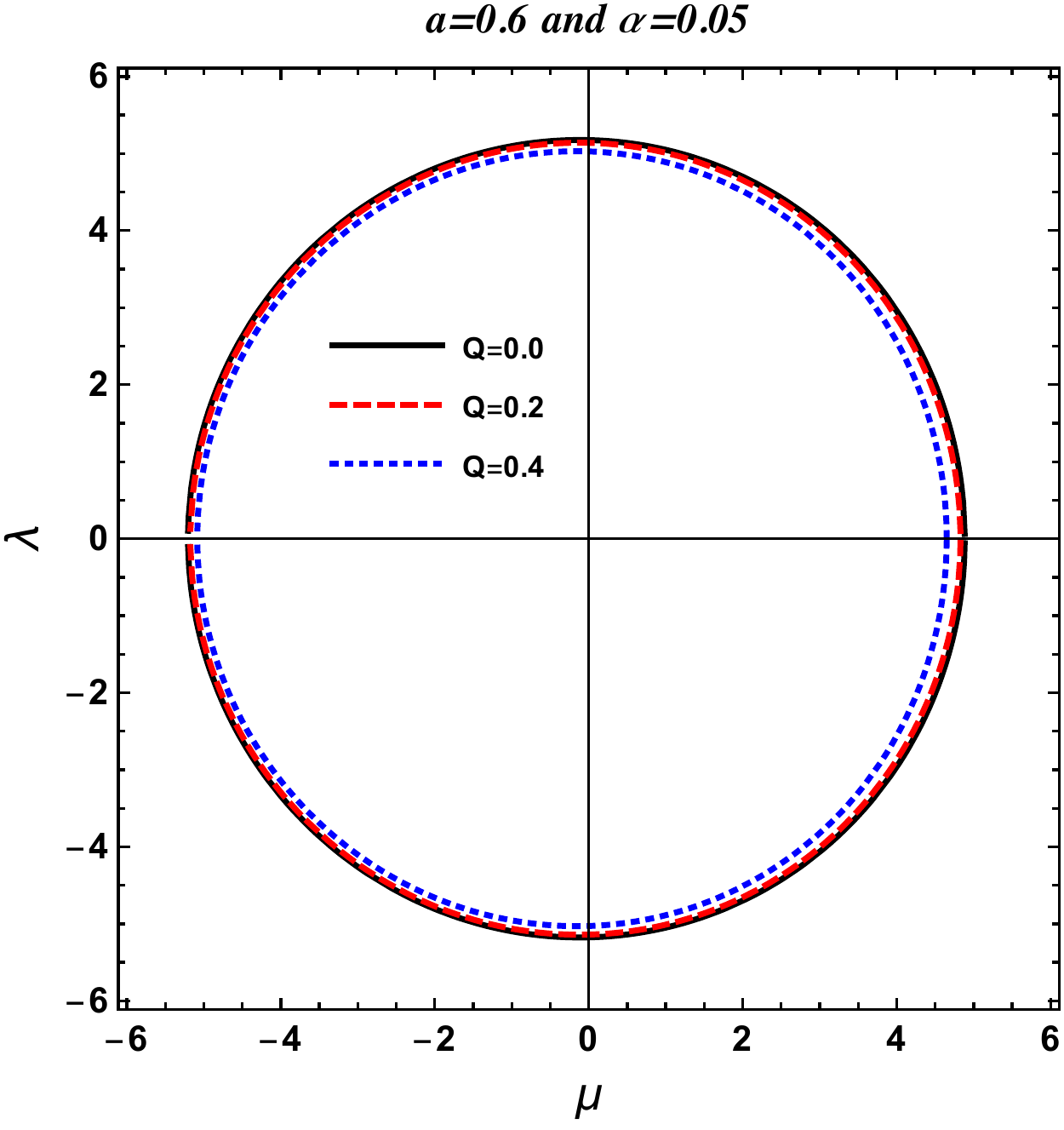}
   \includegraphics[scale=0.45]{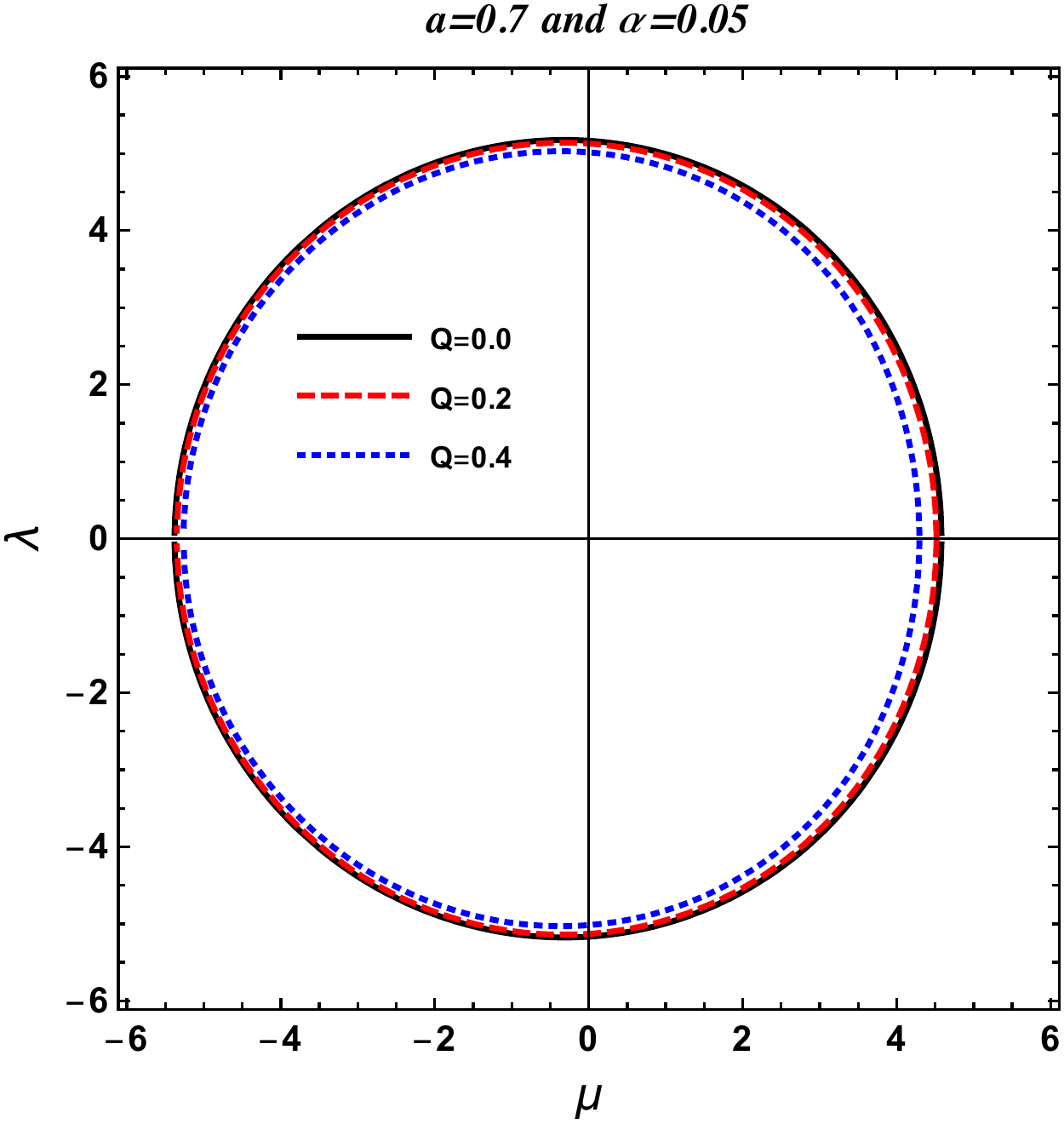}
      \includegraphics[scale=0.45]{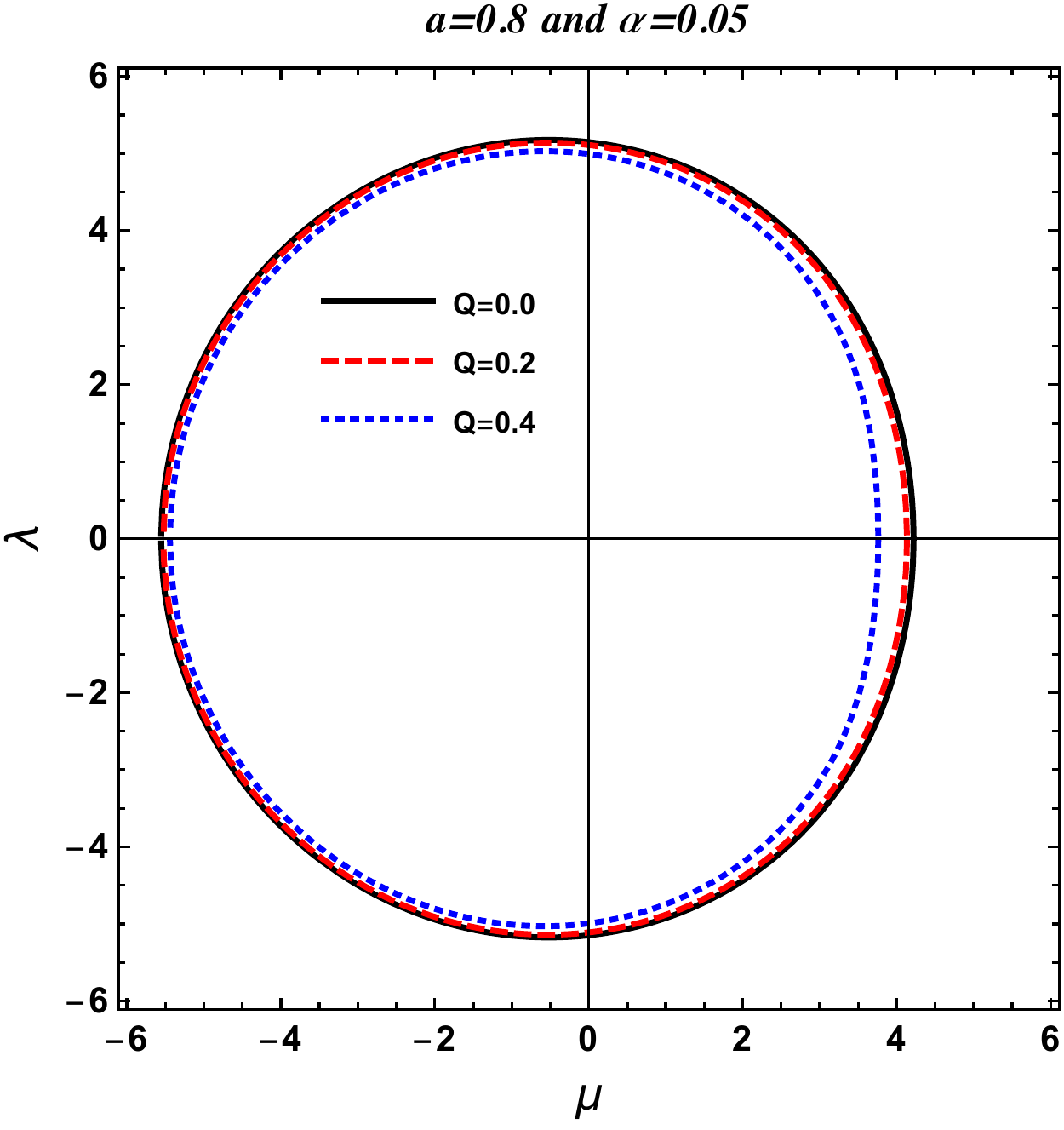}
      
   \includegraphics[scale=0.45]{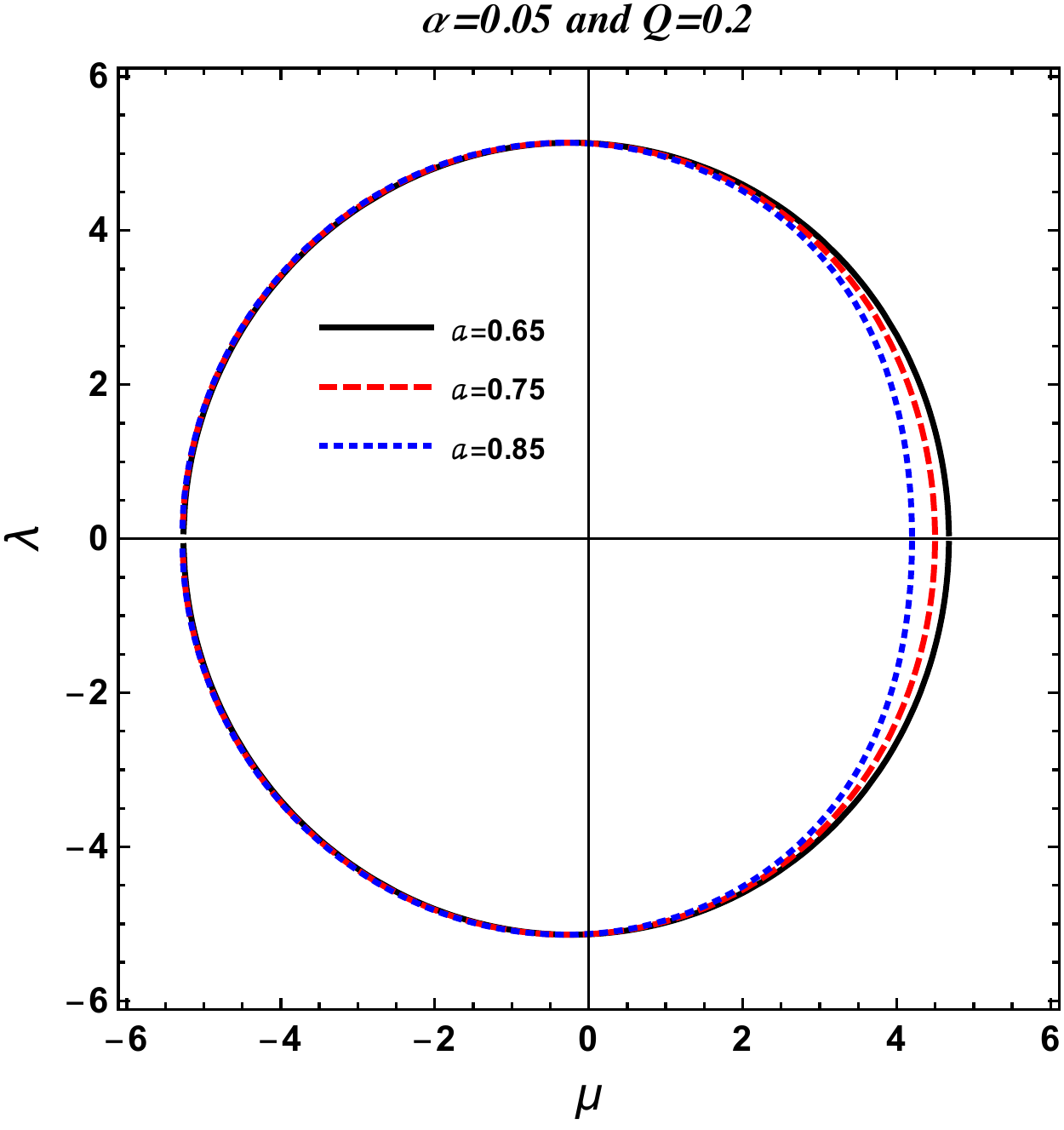}
   \includegraphics[scale=0.45]{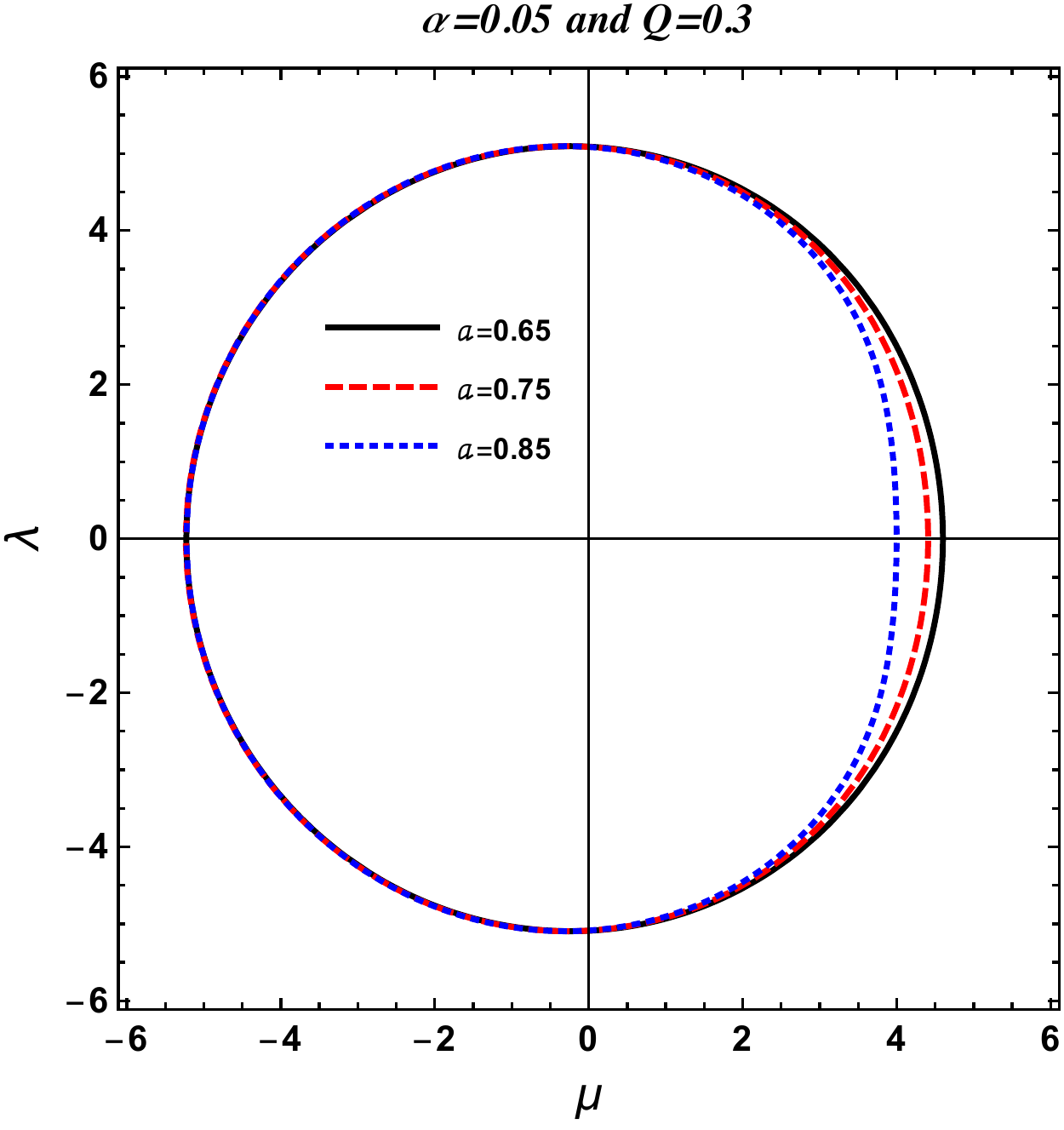}
     \includegraphics[scale=0.45]{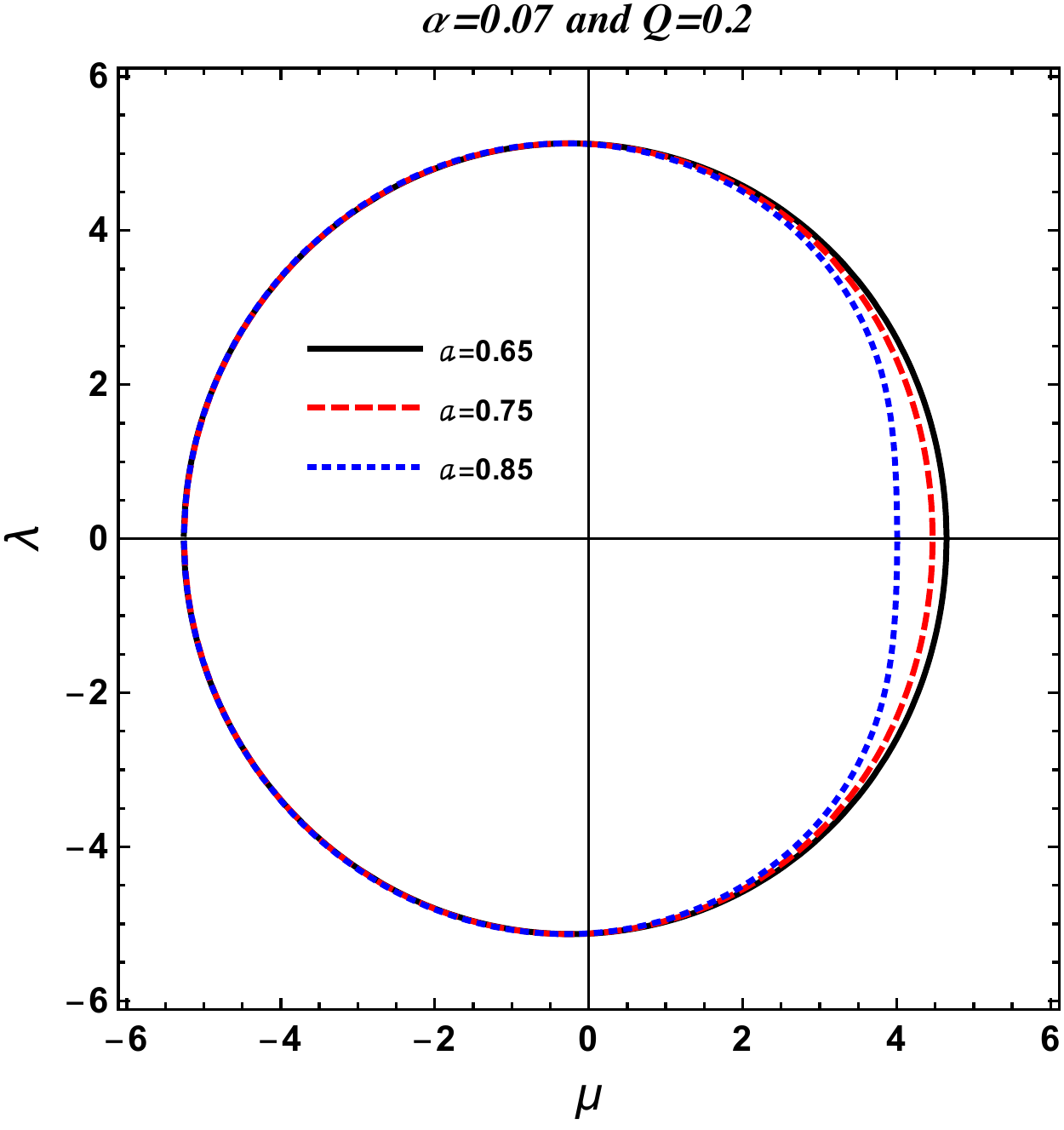}
  \end{center}
\caption{Shadow of the charged rotating BH in 4D-EGB gravity. [Upper panel] For different values of $\alpha$ at fixed $a$ and $Q$. [Middle panel] For different values of $Q$ at fixed $a$ and $\alpha$. [Lower panel] For different values of $a$ at fixed $Q$ and $\alpha$ with $M=1$. }\label{shadowa}
\end{figure*}
\subsection{Rotating black hole shadows}
To determine the geometry of the shadow of charged rotating 4D-EGB BH
we define the celestial coordinates $\mu$ and $\lambda$ as
\begin{eqnarray}
 \mu&=&\lim_{r\rightarrow \infty}
   \bigg(-r^{2}\sin\theta\frac{d\phi}{dr}
      \bigg|_{\theta\rightarrow \theta_{0}}\bigg)
     =-\xi\csc\theta_{0},\label{alpha}\\
 \lambda&=&\lim_{r\rightarrow \infty}
   \bigg(r^{2}\frac{d\theta}{dr}\bigg|_{\theta\rightarrow \theta_{0}}\bigg)
     \nonumber \\ &&=\pm\sqrt{\eta+a^{2}\cos^{2}\theta_{0}-\xi^{2}\cot^{2}\theta_{0}}.\label{beta}
\end{eqnarray}

\begin{figure*}
 \begin{center}
   \includegraphics[scale=0.5]{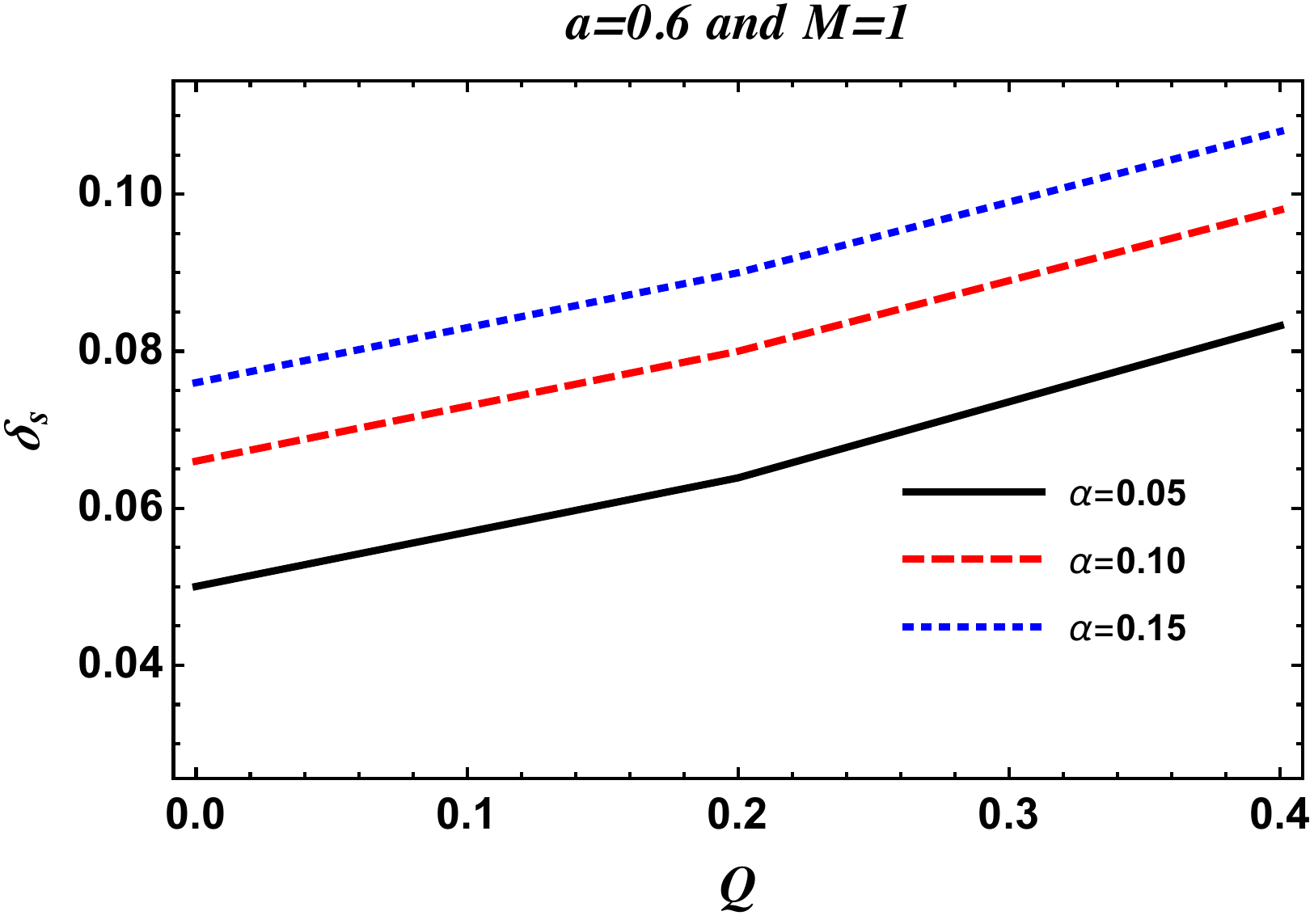}
   \includegraphics[scale=0.5]{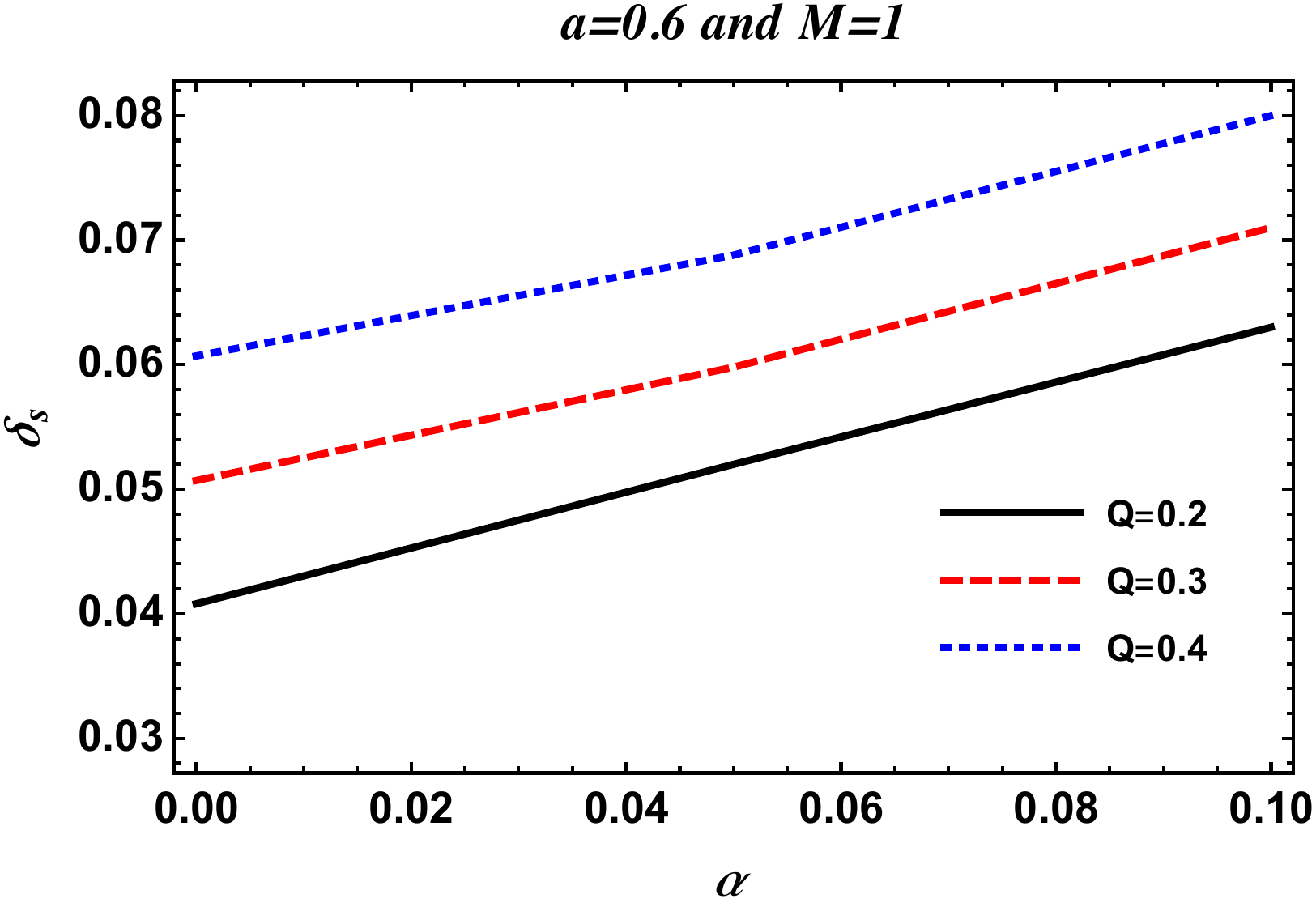}
  \end{center}
\caption{Plot showing the variation of distortion parameter $\delta_s$ with $Q$ for different $\alpha$ (left) and with charge $\alpha$ for different $Q$ (right). Here we have taken $a=0.6$ and $M=1$.  }\label{distortion plot}
\end{figure*}

As we have chosen an equatorial plane for determining the shadow of the BH, so the angle
of the inclination is $\theta_0 = \pi/2$. So, in this case, Eq. (\ref{alpha}) takes the form
\begin{eqnarray}
 \mu&=&=-\xi\\
 \lambda&=&\pm\sqrt{\eta}.
\end{eqnarray}
To visualize the shapes of the shadow of we have plotted the celestial coordinates $\mu$ vs $\lambda$ for different values of the spin parameter $a$, charge $Q$ and GB coupling constant $\alpha$. 
The presence of all the parameters has an influence on the apparent shape and size of the shadow, i.e., with the increase in value of 
$\alpha$, the distortion is increasing at constant $a$ and $Q$. Similarly, for fixed values of $a$ and $\alpha$, increasing $Q$
reduces the size of the shadow and enhances the distortion. However,   the distortion in the shape of the shadow is more in increasing the spin parameter at fixed $\alpha$ and $Q$. 

To further see the effect of all parameters on the shape of the shadow for charged rotating 4D-EGB BH, we have plotted the distortion with respect to $Q$ and $\alpha$ by obtaining it numerically using the shadow plots [Fig. (\ref{shadowa})] in Fig. (\ref{distortion plot}) for different values of $\alpha$ and $Q$. 

\subsection{Emission energy} \label{emission}

It is well known that the quantum fluctuations inside the BHs results in the creation and annihilation of large number of particles near the horizon and the particles with positive energy escape from the BH through tunneling, inside the region where Hawking radiation occurs which causes the BH to evaporate in a certain period of time. Here, we would like to study the associated energy emission rate. Here, the high energy absorption cross section approaches to the BH shadow for a far distant observer. The absorption cross section of the BH oscillates to a limiting constant value $\sigma_{lim}$ at very high energy. It turns out that the limiting constant value is related to the radius of photon sphere as \cite{Wei2013emission,ESLAMPANAH2020115269,Papnoi:2014aaa}
\begin{equation}\label{sigmalast}
\sigma_{lim} \approx \pi R_{sh}^2,
\end{equation}
where $R_{sh}$ is radius of BH shadow, and which provides the energy emission rate expression of BH as \cite{Wei2013emission,ESLAMPANAH2020115269,Papnoi:2014aaa}
\begin{equation}\label{emissionenergyeq}
\frac{d^2 {\cal E}}{d\omega dt}= \frac{2 \pi^2 \sigma_{lim}}{\exp[{\omega/T}]-1}\omega^3,
\end{equation}
where $T=\kappa/2 \pi$ is the Hawking temperature and $\kappa$ is the surface gravity. And combining eq.(\ref{sigmalast}) and eq.(\ref{emissionenergyeq}), we can rewrite a new form of expression for emission energy as below

\begin{equation}\label{emissionenergyeqlast}
\frac{d^2 {\cal E}}{d\omega dt}= \frac{2\pi^3 R_{sh}^2}{e^{\omega/T}-1}\omega^3.
\end{equation}

\begin{figure*}
 \begin{center}
 \includegraphics[scale=0.7]{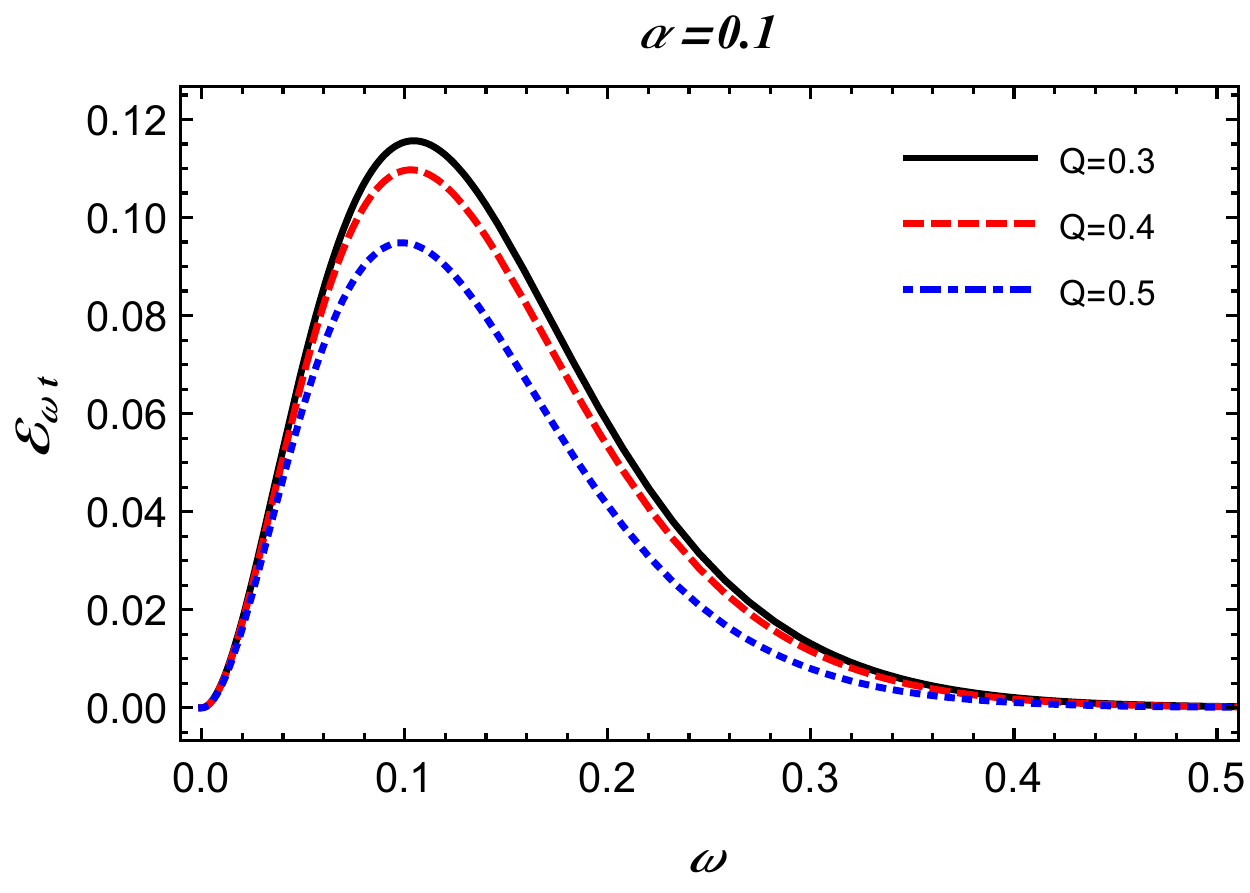}
 \includegraphics[scale=0.7]{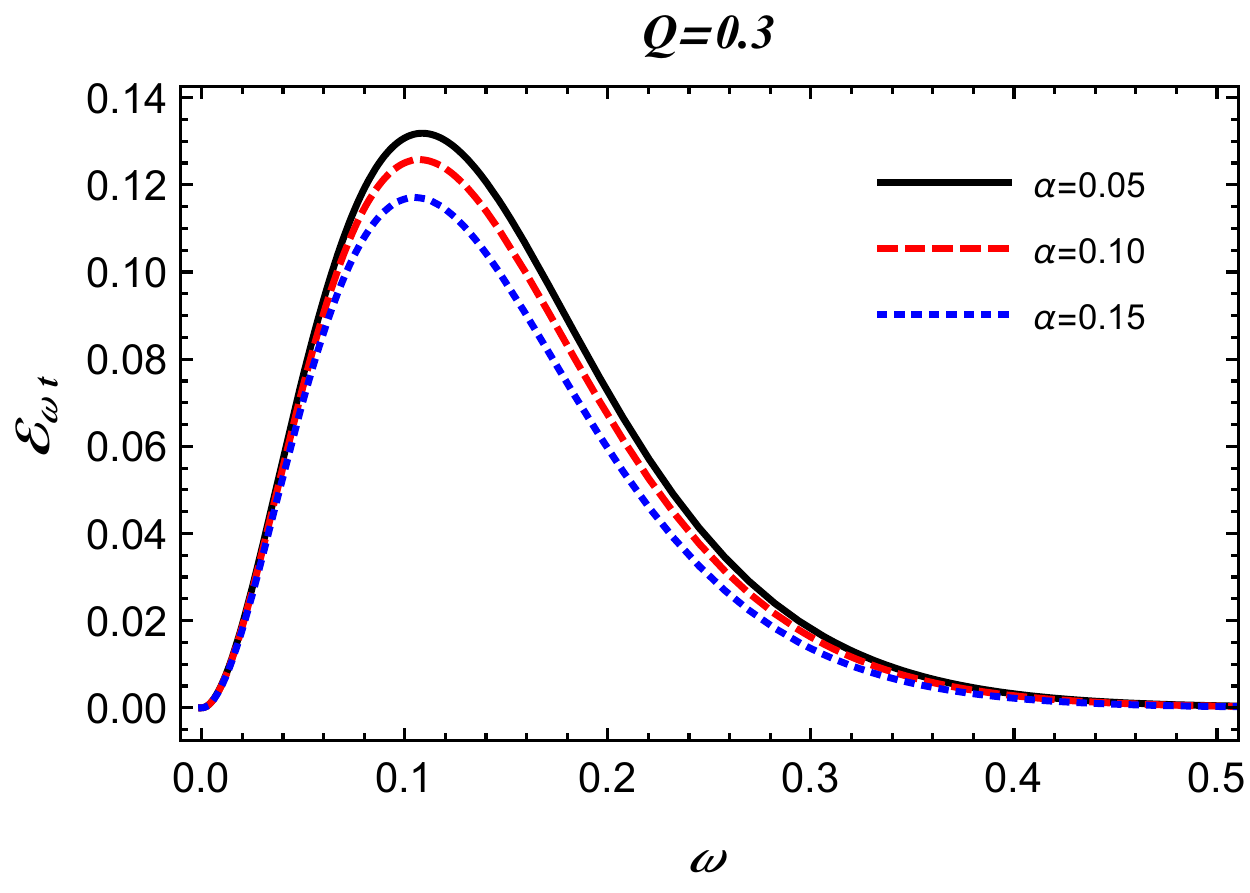}
  \end{center}
\caption{Plots showing the rate of energy emission varying with
the frequency for different values of the GB coupling constant $\alpha$ (Left panel) and  BH charge $Q$ (Right panel).}\label{emission plot}
\end{figure*}

The variation of energy emission with $\omega$ for fixed GB coupling constant at different values of charge and fixed charge at different values of GB coupling constant is represented in Fig.~\ref{emission plot} and it is seen that with the increase in the values of GB parameter and BH charge $Q$, the peak of energy emission rate decreases, which implies that the lower energy emission rate corresponds to the slow evaporation process in BH. For simplicity, we use the following notation as ${\cal E}_{\omega t}=\frac{d^2 {\cal E}}{d\omega dt}$ in Fig.~\ref{emission plot}.

\section{Conclusions and discussions}
\label{Conclusions}
In this paper, we have considered the charged BH in 4D-EGB gravity and obtained its rotating solution via complex coordinate transformation using NJA and hence we have investigated the shape of the shadow for same to see the effect of GB coupling constant on the horizon, effective potential, shape of the shadow and energy emission. With this we obtained the following results:  
\begin{itemize} 

\item{It is found that the GB coupling constant $\alpha$ makes an influence on the horizon radius. In Fig. (\ref{hor}), we have studied the horizon structure for different values of spin parameter $a$, GB parameter $\alpha$ and charge of the BH $Q$. For a fixed value of $a$ and $Q$, there exists an extremal value of $\alpha = \alpha_{\epsilon}$ at which $ r_- = r_+ = r_{\epsilon}$, for $\alpha < \alpha_{\epsilon}$ their exist two distinct horizons and their exist no horizon for $\alpha > \alpha_{\epsilon}$. Similarly, for a given value of $\alpha$ and $Q$, there is an extremal value at $a=a_{\epsilon}$ and for fixed $\alpha$ and $a$ there is an extremal value at $Q = Q_{\epsilon}$ for which $r_- =r_+ =r_{\epsilon}$. It is also seen that the radius decreases with the increase in value of all the parameters.}
\item {We have plotted effective potential with respect to radius in Fig. (\ref{vefflast}) and It is seen that the effective potential exhibit a peak which correspond to unstable circular orbit. It is also seen that peak is increasing and shifting towards left with increase in the value of $a$, $Q$ and $\alpha$ which signifies the shifting of circular orbits to central object.}
\item {We have observed that the size of the shadow in case of non rotating BH decreases with the increase in value of GB parameter $\alpha$, spin parameter $a$ and BHs charge $Q$.}
\item {The effect of $\alpha$ on the shape and size of shadow for charged rotating BH in 4D EGB BH  shows that the distortion increases with the increase in value of $\alpha$ and $a$ and with the increase in value of  $Q$ size decreases and distortion in the shape of the shadow increases.}
\item {The rate of distortion in the shape of the shadow has also been observe by plotting the observable $\delta_s$ with respect to $\alpha$ and $Q$ in Fig. (\ref{distortion plot}) and it can be seen clearly that distortion is increasing with the increase in both the parameters.}
\item {The dependence of the energy emission rate
on the frequency is investigated in Fig. (\ref{emission plot}). 
It is observed that the rate of emission is higher for the small
value of both $\alpha$ and $Q$. Thus, a large amount of energy is liberated at low value of $\alpha$ and $Q$.}

\end{itemize}
All the above mentioned results obtained reduce to the case of usual Kerr-Newman BH, Kerr BH and Schwarzschild BH in GR in the prescribed limit.

\section*{Acknowledgements}
The authors would like to thank the anonymous referee for the instructive comments.
U.P. would like to thank University
Grant Commission (UGC), New Delhi for DSKPDF through
grant No. F.4 a  2/2006(BSR)/P H/18 a 19/0009. U.P. would also like to acknowledge the facilities at ICARD, Gurukul Kangri (Deemed to be University), Haridwar, India. F.A. acknowledges the support of INHA University in Tashkent and this research is partly supported by Research Grant F-FA-2021-510 of the Uzbekistan Ministry for Innovative Development. 
\bibliographystyle{apsrev4-1}
\bibliography{4egb}

\end{document}